\documentclass[final,5p,times]{elsarticle}

\usepackage{amssymb}
\usepackage{amsmath}

\usepackage{soul}
\usepackage{cuted}
\usepackage[dvipsnames,table]{xcolor}
\usepackage[most]{tcolorbox}
\usepackage{helvet}
\usepackage{tabularx}
\usepackage{array}
\usepackage{graphicx}
\usepackage{geometry}
\usepackage[ruled,linesnumbered]{algorithm2e}
\usepackage{subcaption}
\usepackage{adjustbox}
\usepackage[colorlinks=false, hidelinks]{hyperref}
\usepackage[para]{footmisc}
\usepackage{comment}
\usepackage{algorithmic}
\usepackage{tabularx}
\usepackage{booktabs}
\usepackage{microtype}
\usepackage{url}
\usepackage{lipsum}
\usepackage{multicol}
\usepackage{mdframed}
\usepackage{booktabs}
\usepackage{float}

\newboolean{showcomments}
\setboolean{showcomments}{false}
\ifthenelse{\boolean{showcomments}}
{\newcommand{\nb}[2]{
  \fcolorbox{black}{yellow}{\bfseries\sffamily\scriptsize#1}
  {\sf\small\textcolor{teal}{\textit{#2}}}
 }
 
}
{\newcommand{\nb}[2]{}
 
}

\usepackage[normalem]{ulem}
\newboolean{showdiffs}
\setboolean{showdiffs}{false}
\ifthenelse{\boolean{showdiffs}}
{
    
    \newcommand{\ins}[1]{\textcolor{blue}{#1}} 
    \newcommand{\del}[1]{\textcolor{red}{\sout{#1}}} 
}
{
    
    \newcommand{\ins}[1]{#1}
    \newcommand{\del}[1]{}
    
}

\newcommand\pname[1]{\textsc{RobEthiChor}#1}
\newcommand\compname[1]{\textsc{RobEthiComp}#1}
\newcommand\robassist{\textsc{RobAssist}}
\newcommand\roboname[1]{\textsc{RobAssist} \begin{math}#1\end{math}}
\newcommand\appname[1]{\textsc{RobEthiApp}#1}
\newcommand\pnameros[1]{\textsc{RobEthiChor-Ros}#1}

\def\myurl#1{\setbox0\vbox{\hsize.5\maxdimen
\url{#1}\par
\setbox0\lastbox
\global\setbox1\hbox{\unhbox0\unskip\unskip\unpenalty}}\unhbox1 }

\setstcolor{red}

\definecolor{softpurple}{HTML}{BFA2DB}

\journal{Journal of Systems and Software}

\begin{document}

\begin{frontmatter}



\title{
\pname{}: Automated Context-aware Ethics-based Negotiation\\ for Autonomous Robots}


\author[label1]{Mashal Afzal Memon}
\author[label2]{Gianluca Filippone}
\author[label2]{Gian Luca Scoccia}
\author[label1]{Marco Autili}
\author[label2]{Paola Inverardi}

\affiliation[label1]{organization={Department of Information Engineering and Information Sciences and Mathematics, University of L’Aquila},
            addressline={Via Vetoio}, 
            city={L'Aquila},
            postcode={67100}, 
            country={Italy}}

\affiliation[label2]{organization={Department of Computer Science,  Gran Sasso Science Institute},
            addressline={Viale Francesco Crispi}, 
            city={L'Aquila},
            postcode={67100}, 
            country={Italy}}

\begin{abstract}
The presence of autonomous systems is growing at a fast pace and it is impacting many aspects of our lives. Designed to learn and act independently, these systems operate and perform decision-making without human intervention. However, they lack the ability to incorporate users' ethical preferences, which are unique for each individual in society and are required to personalize the decision-making processes. This reduces user trust and prevents autonomous systems from behaving according to the moral beliefs of their end-users.
When multiple systems interact with differing ethical preferences, they must negotiate to reach an agreement that satisfies the ethical beliefs of all the parties involved and adjust their behavior consequently.
To address this challenge, this paper proposes \pname{}, an approach that enables autonomous systems to incorporate user ethical preferences and contextual factors into their decision-making through ethics-based negotiation.
\pname{} features a domain-agnostic reference architecture for designing autonomous systems capable of ethic-based negotiating.
The paper also presents \pnameros{}, an implementation of \pname{} within the Robot Operating System (ROS), which can be deployed on robots to provide them with ethics-based negotiation capabilities.
To evaluate our approach, we deployed \pnameros{} on real robots and ran scenarios where a pair of robots negotiate upon resource contention. Experimental results demonstrate the feasibility and effectiveness of the system in realizing ethics-based negotiation. \pname{} allowed robots to reach an agreement in more than 73\% of the scenarios with an acceptable negotiation time (0.67s on average). Experiments also demonstrate that the negotiation approach implemented in \pname{} is scalable.
\end{abstract}






\end{frontmatter}



\section{Introduction}
\label{sec:intro}

The presence of autonomous decision-making systems is growing at a fast pace and is poised to have a critical impact on society~\cite{suri2023software,jedlickova2024ensuring}.
Increasingly, these systems will be acting on behalf of humans~\cite{anderson2018artificial} and making decisions autonomously without the need for human intervention~\cite{waldman2019power}.
As of today, however, human values and ethics are mostly not considered by the digital systems that are making decisions for them, leading to a lack of 
trust,  
\ins{which in fact requires considering not only technical and legal aspects, but also ethical ones~\cite{lacher2017framework,reinhardt2022trust, floridi2019establishing,kares2023trust}.} 
In particular, the autonomous systems that we have today are not able to consider the ethical preferences of end users, which can indeed be used to customize their decision-making process.
\ins{This inability undermines their 
ethical trustworthiness, which, alongside technical trustworthiness, forms a crucial component of an overall trustworthy system~\cite{donati2025beyond,TOSEM2025}.}
To truly serve our needs and become an integral part of our lives, these systems must expand their abilities to integrate mechanisms that enable ethical considerations in their decision-making process~\cite{inverardi2022ethical,alidoosti2021ethics,anderson2020machine,tolmeijer2020implementations,alidoosti2022incorporating}.
%
This calls for enhancing the decision-making to consider the personal ethical preferences of users and take into account their morals and beliefs, which may vary over contexts and are unique for each individual in society~\cite{alidoosti2022incorporating,alidoosti2025exploring,friedman2013value,awad2018moral,bogosian2017implementation,migliarini2020elicitation}. 
For instance, in fields such as healthcare, where service robots help in patient care, smart home systems, where robots provide personal assistance, autonomous vehicles, etc., it is essential for these systems to take into account the ethical preferences of their users to make ethically-informed decisions. \ins{Engineering autonomous systems that can reflect on ethical considerations will not only enable their integration into society but also foster trust toward these systems~\cite{donati2025beyond,donati2024trust}.}

Although several studies~\cite{liao2019building,liao2023jiminy,dennis2016formal,bremner2019proactive,winfield2014towards,Winfield2019} consider ethics in their decision-making, these systems adhere to ethical principles established by system designers (as detailed below in Section~\ref{sec:related}). Hence, due to the absence of a universal ethic, its intrinsic subjectivity and capillarity~\cite{10.1145/3635715}, the challenge of ``personalizing'' autonomous systems with the moral beliefs of end users (e.g., through an ethical profile) for the purpose of customizing their decision-making 
remains mostly unaddressed~\cite{TOSEM2025}. For instance, how should the system make a decision that accounts for the user's benevolence towards the elderly or people with health problems? 

In this direction, the work in~\cite{cacmInverardi2019} delves into the foundational principles of addressing user ethical concerns in autonomous systems: individuals need to be able to exercise (some degree of) control over the decisions that autonomous systems make on their behalf, i.e., autonomous systems, in their behavior, need to consider user's moral beliefs.
Empowering the user with personalized software connectors that are able to reflect the user's moral values in the interactions with digital systems is also the approach taken in the Exosoul project~\cite{AutiliACCESS19}. The approach has been applied to privacy profiles~\cite{DiRuscioIMN24,InverardiMP23} and ethical profiles~\cite{AlfieriHHAI22}.
Following this line of work, in~\cite{boltz2024human}, the authors emphasize the need for user empowerment within the framework of autonomous systems, which demand mechanisms for balancing diverse needs, values, and ethics at the individual, community, and societal levels.

An important practical consideration here is that, similarly to what often happens to humans, autonomous systems are not completely ``isolated''; rather, they operate in a shared physical environment, where they have to deal with each other, e.g., to solve situations where resources are contented, such as using the only one available elevator or tool, occupying a parking spot, traversing a narrow corridor/door as first~\cite{palmer2018modelling,youssefmir1995resource,brambilla2008coordination,nam2015assignment,stavrou218optimizing,jiang2019multi}.
Thus, systems may need to interact to reach a situational agreement that satisfies the moral beliefs of the users they are acting on behalf of~\cite{AlfieriDGGS23,AlfieriHHAI22,bogosian2017implementation,bostrom2018ethics,nallur2019ethics,ryan2020artificial}. The agreement is situational in that it is unrealistic to expect that it is reached once and for all; rather, it is situational in that it may depend on the specific context and current status of the users.
For example, how can an assistive robot decide whether to give priority to another assistive robot, considering the status of the users being assisted if they are injured, elderly, or in a hurry when in the supermarket or the hospital?
Negotiation is a possible way of reaching such an agreement~\cite{chen2019automated,khemakhem2020agent} and \emph{automated negotiation} is the process through which multiple autonomous systems communicate by automatically exchanging bids, dialogues, and offers for that purpose~\cite{memon2025systematic,khemakhem2020agent,baarslag2017automated,kiruthika2020lifecycle,bagga2022deep,zuckerman2013towards,baarslag2016learning,memon2023automated}.
However, quantifying user ethical preferences and formalizing negotiation rules based on them remains another key challenge~\cite{chen2019automated,kraus2011agents}.

To address these challenges, this paper presents \pname{}, an approach that makes autonomous systems capable of exploiting user ethical preferences to regulate their autonomous behavior and support decision-making through ethics-based negotiation.
\ins{The \pname{} approach aims to contribute towards the design and development of trustworthy autonomous systems capable of replicating the behavior of humans while making ethically informed decisions on their behalf. The goal is to promote the system's trustworthiness by adopting an architecture that, supporting automated ethical negotiation, allows systems to take into account ethical considerations at runtime~\cite{yazdanpanah2021responsibility,abeywickrama2023specifying,DBLP:conf/icse/BennaceurHNZ23}.} Specifically, we target autonomous systems that act on behalf of humans~\cite{anderson2018artificial} and, building on the principles of Utilitarianism~\cite{moor2006nature,mill2016utilitarianism,allen2005artificial}, exploit a user's ethical profile and a context model to personalize their autonomy and, upon a resource contention, negotiate to reach an agreement. The approach features a domain-agnostic reference architecture as a template solution for realizing autonomous systems capable of leveraging the ethical profiles of end users and their contextual status during negotiation, so to ensure that the decision is aligned with the preferences of all parties involved.
In this paper, we also present \pnameros{}, a specific instance of our reference architecture implementing the proposed automated context-aware ethics-based negotiation on top of the Robot Operating System (ROS). 
The full implementation of  \pnameros{} is available\footnote{https://github.com/gianlucafilippone/robethichor-ros}, and the experimental results show that the approach effectively accomplishes the negotiation and allows systems to behave according to user ethical preferences. The replication package permits reproducing the experiments\footnote{https://doi.org/10.5281/zenodo.17449923}. In summary, main contributions are:

\begin{itemize}
    \item[$\bullet$] a context-dependent ethical profile for specifying user ethical preferences in terms of graded dispositions;

    \item[$\bullet$] an automated context-aware ethics-based negotiation process for ethically-informed decision making;

    \item[$\bullet$] a domain-agnostic reference architecture providing a template solution for realizing autonomous systems supporting ethics-based negotiation;

    \item[$\bullet$] \pnameros{}: the full implementation of the architecture for the robotic domain through the Robot Operating System (ROS);

    \item[$\bullet$] the evaluation of \pname{} to assess its effectiveness in realizing the negotiation and its viability with real robots.

\end{itemize}

The rest of the paper is organized as follows. Section~\ref{sec:background} introduces the main ingredients of our approach. Section~\ref{sec:scenario} presents a motivating scenario, and Section~\ref{sec:approach} details the proposed solution. Section~\ref{sec:architecture} describes the reference architecture, and Section~\ref{sec:implementation} presents its implementation. Section~\ref{sec:evaluation} presents the evaluation of the approach and its results. Sections~\ref{sec:discussion} and~\ref{sec:threats} discuss relevant aspects of our approach and highlight limitations and threats to validity. Section~\ref{sec:related} discusses related work, and Section~\ref{sec:conclusion} concludes the paper with future steps.

\section{Background and scope}
\label{sec:background}

In the following, we present an overview of the main ingredients/basic concepts that we leverage in our work.

\subsection{Digital ethics}
We employ the notion of \textit{digital ethics}, as proposed by Floridi 
\cite{floridi2018soft}, which differentiates between \textit{hard ethics}, referring to the ethical rules described by the higher authorities which are, mostly often, domain-specific and, in principle, commonly accepted, and \textit{soft ethics}, which encompasses the individual's ethical preferences~\cite{floridi2018soft}, being individuals single people, groups, associations, etc. We consider that autonomous systems comply with the hard ethics of the domain/context and negotiate based on their users' soft ethical preferences to (possibly) reach an ethical agreement that aligns with the ethical preferences of all parties involved.
The user's soft ethical preferences that we consider in this paper are inspired by the notion of \textit{dispositions}~\cite{Anjum_dispositions:2013}.
In its most general philosophical acceptation, dispositions refer to the capacity of an object to behave differently under different circumstances~\cite{spehrs2024dispositional}.
For example, water exhibits different behaviors according to environmental conditions, i.e., it becomes solid at freezing temperatures, liquid under moderate temperatures, and vaporizes upon heating, adapting its form naturally to the environment.
Similarly, a given user may act differently in a hospital compared to an airport based on his/her contextual conditions/status/dispositions (as we will describe in Section~\ref{sec:scenario}).
Dispositions are gradable properties, associated with the context, which can be instantiated by any kind of entity~\cite{donati2024representing}.
\textit{Grades} are assigned to dispositional properties according to the importance that the users attribute to each disposition, i.e., the willingness of a user to behave in a certain way under specific circumstances, thereby reflecting their preferences/values within a specific context.
More recently, the problem of how to embed human values in the design and implementation of software systems, notably the problem of how to operationalize human values, has received increasing attention in the software engineering community \cite{AutiliACCESS19,DBLP:conf/sigsoft/MougoueiPHSW18,shahin2022operationalizing, DBLP:conf/icse/BennaceurHNZ23}. Beyond our approach to modeling ethical preferences through dispositions, all those approaches can be well exploited to provide user ethical profiles that enjoy similar properties.

\subsection{Context awareness}
We utilize Dey's~\cite{dey2001understanding} notion of context and context awareness for building our work. \textit{Context} is defined as any information that can be used to characterize the situation of an entity, i.e., person, place, object, and \textit{context awareness} is described as the system's ability to utilize contextual information in order to dynamically adapt according to the user's needs. The system requires context management facilities to effectively incorporate contextual information, allowing it to adjust/customize its autonomy in real-time. In our work, \textit{context} consists of two primary components: environmental factors such as location, time, and other pertinent elements, and \textit{user status} which is further defined as a set of physical, social, or emotional conditions that (possibly) apply to the user in a given context, e.g., elderly, injured, and crowd anxiety~\cite{dey2001understanding,xu2015cina}.
The user status influences the autonomous system's decision taken on the user's behalf as it provokes the activation of dispositions within a given context.

\subsection{Context-dependent ethical profile}\label{sub:ethical-profile}
The integration of the above-mentioned concepts forms the cornerstone of our \textit{ethical profile}. The ethical profile is defined as a set of context-dependent dispositions graded by the user to specify their preferences in a given context which are further utilized by autonomous systems to adjust their behavior. As detailed in Section~\ref{sub:ethical-impact}, the dispositions then lay the foundation for calculating the ethical implications and the ethical impact of the system decisions.
This means that, according to their soft ethics, users can customize the ethics-based behavior of the systems that will act on their behalf by specifying the grades for the dispositions (within the profile) they care about. For that purpose, we build on the intuition in~\cite{dennis2016formal,Machine_Ethics_in_Changing_Contexts:2021}, the architectural layer proposed in~\cite{bremner2019proactive,VANDERELST201856}, and the findings in~\cite{Nallur:2020,tolmeijer2020implementations,alidoosti2022incorporating,alidoosti2025exploring}, according to which (i) moral preferences are considered soft constraints, rather than hard vetoes on tasks; (ii) explicit operationalization of ethical concepts require some encoding of values to perform utility-based reasoning; (iii) changes of context can affect the encoded values; (iv) systems can come with predefined, yet adjustable~\cite{mostafa2019adjustable}, domain-specific moral preferences together with their default values. 
In particular, as detailed in Section~\ref{sec:related}, the work in~\cite{dennis2016formal,Machine_Ethics_in_Changing_Contexts:2021,alidoosti2025exploring} inspired the operationalization of ethical values by associating context-dependent grades to ethical principles, hence permitting the calculation of the ethical utility, and the work in~\cite{alidoosti2022incorporating,Nallur:2020,tolmeijer2020implementations} inspired the incorporation of the profile into our architecture, in turn inspired by architectural layer proposed in~\cite{bremner2019proactive,VANDERELST201856} for enabling ethical reasoning on robots.

It is worth clarifying that the main focus of our work does not concern the elicitation, modeling, and verification of ethical rules as in~\cite{feng2023towards, sinem2025specification}. Similarly, it does not focus on the generation of ethical profiles, which encapsulate users' context-specific ethical preferences; rather, it concerns their usage, and hence on how they can be integrated into the systems to support decision-making based on users' ethical preferences. As already anticipated in the introduction, there are various approaches in the literature for constructing such profiles using questionnaires, surveys, product reviews, or social media interactions, as in ~\cite{AlfieriHHAI22,TOSEM2025,cacmInverardi2019,InverardiMP23,DiRuscioIMN24} and references therein.

\subsection{Resource contention}\label{sub:resource-contention}
\textit{Resource contention} refers to those application-specific situations in which two or more autonomous systems are interested in doing something, but not all of them can do it simultaneously~\cite{palmer2018modelling,youssefmir1995resource,brambilla2008coordination,nam2015assignment,stavrou218optimizing,jiang2019multi}. This may happen because the resource can only be mutually exclusively accessed, used, occupied, or consumed. For instance, two robots cannot traverse a narrow corridor or use the same tool simultaneously, dock at the same charging station, and, in general, occupy the same physical space.
An important consideration is that, independently from the negotiation-based approach we are proposing in this work, autonomous systems are often natively capable of using sensors to recognize situations such as the ones above~\cite{pappas2024adaptive,shek2021current,padmasiri2020automated}. That is, in our implementation, we use the built-in sensing capability to detect resource contentions and, hence, the situations where negotiation is needed.

\subsection{Goal-Task-Action} \label{sub:goal-task-action}
We adopt the \textit{Goal-Task-Action} model, which is considered to be a flexible model as most controller architectures can be transformed into it~\cite{VANDERELST201856}. 
The goal is what the system intends to do (e.g., \textit{transporting the passengers at the airport to their departure gates using wheelchairs}); goals are achieved through the accomplishment of tasks (e.g., $t_1= $ \textit{identify the passenger}, $t_2= $ \textit{navigate the path to departure gate}, $t_3= $  \textit{locate the nearest elevator},  $t_4= $  \textit{take the elevator}, $t_5= $ \textit{reach the departure gate}); tasks are decomposed into a set of actions, i.e., elementary operations which can be performed by the system in the environment to achieve the goal (e.g., $a_1 = $ \textit{push the wheelchair towards the elevator}, $a_2= $ \textit{raise the arm to activate the elevator call button}). 
Now, it is worth clarifying that, once tasks are available, a traditional autonomous system controller (with no support for ethics), immediately performs the related actions; our controller, when (and only when) a resource contention is detected, starts the negotiation.
Thus, the main issue we aim to address is not task allocation, in that we assume tasks that have already been assigned to robots as in any traditional controller; 
rather, we focus on the association of ethical values with robot behaviors that are understandable to the users they are acting on behalf of, and on the negotiation over contended resources.
For this reason, and specifically for the purpose of ethics-based negotiation to decide who gets to use a shared resource, we chose to work with tasks (e.g., take the elevator) rather than elementary actions (e.g., raise the arm and press the button), as tasks are higher-level constructs that are more meaningful and relevant to the end user. That said, technically speaking, for the purpose of negotiation, whether we consider tasks or actions does not fundamentally change the approach. The real challenge lies in negotiating ethics itself using an automated negotiation framework, an approach that, in the literature, has been primarily applied to economic bargaining~\cite{memon2025systematic}.

\section{Motivating scenario}
\label{sec:scenario}

\begin{figure*}[ht!]
\centering
\includegraphics[width=0.9\textwidth]{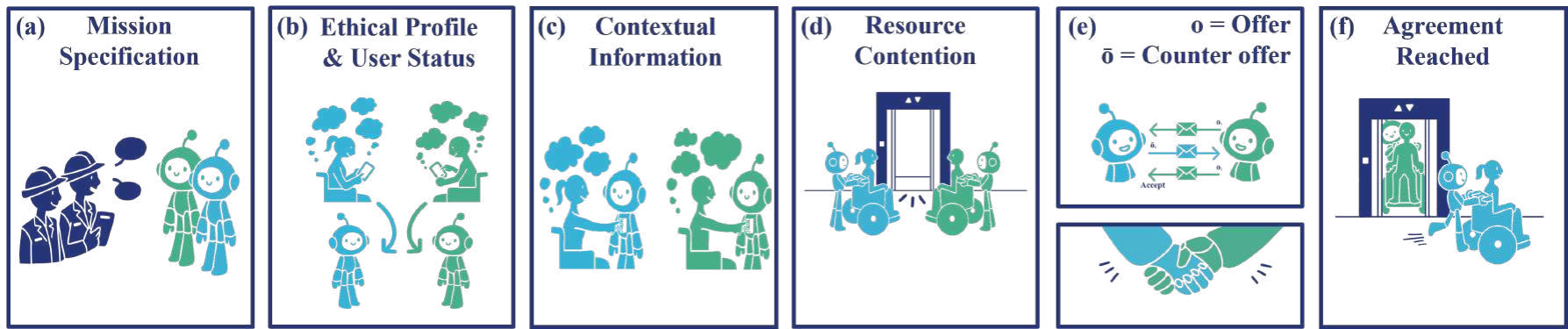}
\caption{Negotiation example for passenger assistance}
\label{fig:roboexample}
\end{figure*}

Industry reports and data from the world's major airports reveal that an average of between 1.2 and 2.4\% of passengers ask for a Special Service Request (SSR). 54\% of those passengers need wheelchair assistance from the entrance to the departure gate, due to a series of factors like disability, long walking distances, and help with baggage. Notably, the trend of wheelchair requests is expected to increase by 15-20\% per year, with major airports already experiencing thousands of requests daily~\cite{urbanrobotics2023, guardian2023heathrow, budd2020supporting}. This trend highlights the need for support systems, which robotics technologies can provide in a wide variety of contexts, such as hospitals, transportation hubs, retail centers, museums, etc~\cite{urbanrobotics2023}. Research has explored the use of autonomous robotic wheelchairs and wheelchair towing systems to provide mobility support, reduce the workload on human personnel, and enhance users' independence~\cite{Baltazar2021autonomous, morales2013human, hsu2012mobility}. These considerations led us to design a motivating scenario where service robot-guided wheelchairs provide smart mobility assistance.

\subsection{Setting}
In this general context, we envision a scenario where a fleet of autonomous assistive robots, named \robassist{}, operates to facilitate users for multiple tasks in different contexts, e.g., in hospitals, airports, etc.
\robassist{} robots are developed by the (imaginary) tech company \compname{} and adopt the \pname{} architecture.
\robassist{}(s) can be customized for various partners/operators/consumers for missions aimed at helping people, e.g., providing physical assistance, navigation assistance, communication support, etc. 
An accompanying app \appname{} enables end users to specify their digital ethical profile through a set of dispositions and associated grades configured for different contexts, which are subsequently utilized by the robots to adjust their autonomy for the relevant context. Importantly, the profile does not need to be generated every time. Once created, it can be reused across different application scenarios, provided that the applications adopt the \pname{} architecture (as in the following scenario, where this is the case for those offered by the tech company \compname{}).
%
Motivated by the airport passengers assistance needs described above, we consider an airport scenario in which multiple robots, alongside human employees, are responsible for a multitude of tasks required to carry out the airport operations, e.g., security screening, customer services, luggage handling, and boarding assistance. Each robot has a dedicated role and a mission according to its capabilities. The robots possess the ability to identify resource contention and, when needed, devise an alternative plan to accomplish their missions if the initial plan becomes unsuitable or unachievable. We consider two assistive robots, \roboname{A} and \roboname{B}, designed by the \compname{} company to navigate in the 
airport environment and offer mobility assistance services to passengers, such as transporting them to their gates, restrooms, and other airport facilities.

\subsection{Scenario description} \label{sub:scenario}

A visual representation of the scenario is presented in Figure~\ref{fig:roboexample}. \roboname{A} and \roboname{B} are programmed with the mission of transporting passengers to their gates using service wheelchairs (Figure~\ref{fig:roboexample}a).
%
$Alice$ is an athlete who suffered a leg injury and hence would prefer assistance to move through the airport.
$Bob$ is an elderly traveler and is running late for the flight, which is boarding shortly. $Bob$ requires assistance in rapidly moving to the departure gate.
The two assistive robots \roboname{A} and \roboname{B} are assigned respectively to $Alice$ and $Bob$ to provide them the required assistance.
%
When encountering \roboname{A} and \roboname{B}, $Alice$ and $Bob$ use their mobile phones to scan the robots and provide their digital ethical profiles through the app \appname{}.
Additionally, they use the app to inform the assistive robots of their current status by selecting a set of possible conditions that apply to them, e.g., elderly, injured, etc. (Figure~\ref{fig:roboexample}b).\footnote{Note that, using the companion app, profiles can be defined beforehand and reused for different scenarios and any compatible system. Similarly, user status conditions can be provided to the app beforehand and updated on the fly, if desired.}
%
$Alice$ and $Bob$ further provide their flight details by scanning the boarding cards, hence allowing the robots to dynamically identify other relevant contextual information, such as flight status updates, etc (Figure~\ref{fig:roboexample}c). 

\subsubsection{Contextual factors}
The overall ethical profile within the \appname{} includes dispositions configured for all contexts managed by the central \robassist{} service offered by \compname{}. 
%
However, in the current location, only the subset of airport-related dispositions are relevant and tailored to additional contextual factors, e.g., the approaching of the flight time, the crowd at the airport, and the change of the boarding gate.
This enables the assistive robot to select the appropriate (sub) profile based on the current context and adjust its behavior according to the grades given by the users to the subset of dispositions currently relevant, further accounting for the status conditions of the users.

\subsubsection{Resource contention}
The airport elevators can only accommodate one passenger in a wheelchair, together with the accompanying robot, at a time. Thus, in busy situations, the usage of an elevator may constitute a \textit{resource contention} and the robots collaboratively negotiate with each other to decide which one will be the first to use the elevator, based on their users' ethical preferences and current status (Figure~\ref{fig:roboexample}d and~\ref{fig:roboexample}e). 
The robots employ a minimalist strategy during the negotiation in order to disclose only the strictly necessary information from the passengers' personal information during the negotiation\footnote{It is worth saying that, although concerns about privacy and security are out of scope, when the target is a real production environment, the approach must be designed so to achieve robustness and security by exploiting state of the art solutions. When the negotiation process ends and the mission is complete, the robots destroy passengers' information.}~\cite{hajaj2017enhancing}.

\subsubsection{Negotiation}\label{sub:scenario-negotiation}

\roboname{B}, while assisting $Bob$, generates an initial offer $o_1^B$ based on $Bob$'s elderly status, suggesting that it should have priority access to the elevator.
%
The offer $o_1^B$ is forwarded to \roboname{A}. When received and evaluated by \roboname{A}, the offer $o_1^B$ is rejected, as $Alice$'s ethical profile indicates a stronger tendency to prioritize self-care due to the injury over giving precedence to an elderly person. 
As a result, \roboname{A} generates an alternative counter offer $o_1^A$ disclosing $Alice$'s injured status. 
When \roboname{B} receives the offer $o_1^A$ from \roboname{A}, it in turn immediately rejects the offer $o_1^A$ since \roboname{B}, monitoring $Bob$'s flight itinerary, detected a gate change for the scheduled flight that is boarding shortly. 
Consequently, \roboname{B} generates a counter offer $o_2^B$, further emphasizing $Bob$'s emergency of scheduled departure and the urgent change in the itinerary. The offer $o_2^B$ is forwarded to \roboname{A}. Upon evaluation, the offer $o_2^B$ of \roboname{B} is accepted by \roboname{A} as $Alice$'s ethical profile indicates a strong disposition towards giving precedence to others in an emergency that, together with the disposition towards giving precedence to the elderly, overrides the priority of self-care.
Upon agreement, $Bob$ and \roboname{B} use the elevator first to ensure arriving at the boarding gate on time while $Alice$ and \roboname{A} wait for the next opportunity to use the elevator (Figure~\ref{fig:roboexample}f). 

An important observation here is that the agreement has been reached collaboratively and is situational in that the decision made is contextually ethical. In fact, given the current status of the users, in the current location, and at the considered time, \roboname{A} assisting Alice willingly gives up and acknowledges the virtuosity of the received offer due to Alice's benevolent disposition to give precedence to elderly when in an emergency. Should, instead, the negotiation end with no agreement, a fall-back strategy that resorts to the hard ethics is employed for the decision-making (e.g., first-come-first-served).

Moreover, one could argue that Alice and Bob could, in principle, directly negotiate without ``delegating'' the decision to the autonomous robots. However, it could be the case that, due to specific conditions (not explicitly mentioned in our proof-of-concept scenario so to keep it ``lighter'', e.g., different languages, particular status conditions, privacy or shyness, and in general any inability), Alice and Bob cannot, do not want, or prefer not to communicate directly. In general, there are situations, like in self-driving cars~\cite{AutiliACCESS19}, where users cannot communicate directly, being in different vehicles. Still, another well-fitting case is that of avatar robots~\cite{Avatarwork:2020,low-moral-avatar-rob:2025} for remote (tele)presence or apparent participation in distant events that, in the near future, will be more and more autonomous, hence enabling human abilities in remote places by also considering ethical aspects. Nonetheless, we utilize the concept of adjustable autonomy~\cite{mostafa2019adjustable}, enabling the system to operate with reduced autonomy and, when possible, transfer control to users if required. This could be the case of Alice and Bob regaining full control and directly negotiating.
Moreover, there can also be situations where the negotiation does not take place or can be overruled, at any stage independently from any possible outcome, due to, e.g., system requirements, organizational aspects, emergency events that
are handled by the behavior programmed into the native robot controller by default.

\subsection{Negotiating in a different context}\label{sub:negotiating-context}
Considering the same needs for assisted mobility mentioned above, and following the \textit{Welcoming people to the hospital} example described in RoboMAX~\cite{robomax}, we consider a second scenario where autonomous robots welcome visitors and patients in a hospital and, if requested, accompany them to their destination.
As in the previous scenario, $Alice$ and $Bob$ both require medical care: $Alice$ is suffering a leg injury, and $Bob$ is an elderly person.
Both prefer to avoid prolonged walking, so they require assistance for moving to the dedicated departments for their medical treatment.
Now, $Alice$ and $Bob$ are assisted by \robassist{}(s) $A$ and $B$ to help them navigate through various hospital departments using wheelchairs. As in the example scenarios described in~\cite{jiang2019multi,wong2015lets}, the robots find themselves in a conflictual situation when they need to traverse a narrow corridor in opposite directions.
Differently from the airport context previously described, now, $Bob$'s ethical profile reflects $Bob$'s disposition to give precedence to the medical needs of an injured person over personal treatment needs.
Consequently, now, \roboname{A} and $Alice$ proceeds through the corridor, while $Bob$ either waits for $Alice$ to pass through it or finds another path. 
These simple scenarios demonstrate that the same users, in different contexts, can have different preferences and exhibit different behaviors. As a result, the outcome of the negotiation can vary. 
In addition, other situational parameters (e.g., delayed treatment, sudden symptoms, waiting in reduced oxygen environments, being in an area with infection risk, etc.) within the hospital context could also lead to a varied negotiation outcome.

\section{Approach description}
\label{sec:approach}

In this section, we give a lightweight formalization that permits us to define: (i) the representation of the context, the user status, the user ethical profile, and the ethical impact of the autonomous system's behavior, (ii) the utility function and decision-making we use to automate the ethics-based context-aware negotiation process. The same formalization will also serve to ease the description of the reference architecture (Section~\ref{sec:architecture}), its instantiation and implementation for ROS (Section~\ref{sec:implementation}), and the description of the experiment we have conducted (Section~\ref{sec:evaluation}).

\subsection{Context model}
When dealing with ethics, the notion of context is a broad concept that encompasses not only simple dimensions such as location, time, weather conditions, and proximity in space (and in general any situation that can be sensed, detected, or recognized), but it can also involve information about the (current) \textit{user status}, e.g., physical, social, emotional state, intent, and sentiment~\cite{Machine_Ethics_in_Changing_Contexts:2021,bauer2017consolidated,dey2001understanding}. 
Many techniques exist for characterizing and reasoning with contexts~\cite{hong2009context,hoyos2013domain,bettini2010survey,caesar2023process,filippone2022synthesis}. 
We adopt a \textit{key-value} approach. 

A context is a set of pairs $\mathcal{C} = \{ \langle a_1, v_1 \rangle, \cdots, \langle a_n, $ $ v_n \rangle \}$, $n$$>$$0$, where each $\langle a_i, v_i \rangle$, $1$$\le$$i$$\le$$n$, is a pair associating the context attribute name $a_i$ with its value $v_i$. Attribute names are keywords that, by default, are associated with the type of the attribute, e.g., boolean, string, integer, address/coordinates, enumerated.

For instance, 
$\mathcal{C}^A_a = \{ \langle location,$ $ airport \rangle,$ $\langle flight\_time,$ $ 20$$:$$30 \rangle,$ $\langle flight\_gate,$ $ 11 \rangle \}$
and 
$\mathcal{C}^B_a = \{ \langle location,$ $ airport \rangle,$ $\langle flight\_time,$ $ 19$$:$$00 \rangle,$ $\langle flight\_gate,$ $ 42 \rangle \}$ 
model the two contexts considered in the motivating scenario in Section~\ref{sec:scenario} for $Alice$ and $Bob$, respectively. Similarly,
$\mathcal{C}^A_h = \{ \langle location, hospital \rangle$, $\langle appointment\_slot,$ $09$$:$$00 \rangle,$ $\langle care\_department,$ $orthopedics \rangle \}$ 
and
$\mathcal{C}^B_h = \{ \langle location,\\hospital \rangle$, $\langle appointment$ $\_slot,$ $09$$:$$00 \rangle,$ $\langle care\_department,$\\$penumology \rangle \}$
model the contexts in the hospital. Note that the definition above does not model nested collections of attribute-value pairs even though, in our implementation, we use JSON\footnote{https://www.json.org}, which supports arbitrarily deep nesting. Its extension is straightforward.
In Section~\ref{sec:discussion}, we point out some of the limitations of the key-value approach (even though empowered with nested pairs) and discuss how more powerful context modeling techniques could be profitably applied.

As part of the context, we also make use of the mentioned user status that, at negotiation time, can affect the result of the agreement being possibly reached. However, we define the user status as a separate model since it has a different scope (i.e., it concerns a subjective dimension that is strictly bound to the user) and a different role in the negotiation process.

\subsection{User status model}
Similarly to the context model above, user status is modeled as a 
set of pairs $\mathcal{S} = \{ \langle c_1, v_1 \rangle, \cdots, \langle c_m, v_m \rangle \}$, $m$$>$$0$, where each $\langle c_j, v_j \rangle$, $1$$\le$$j$$\le$$m$, is a pair associating the name of the user condition $c_j$ with its value $v_j$$\in$$\{true,false\}$. Thus, differently from the context model, the type of the pairs is always boolean. 


\begin{table*}[tph]
    \centering
    \small
    
    \begin{tabular}{c|cc|cc|}
\cline{2-5}
& \multicolumn{2}{|c|}{\textbf{Ethical profile}} & \multicolumn{2}{c|}{\textbf{Alice's grades}}                                              \\ 
\hline
\multicolumn{1}{|c|}{\textbf{User status conditions}} & \multicolumn{1}{c|}{\textbf{Disposition ID}} & \textbf{Disposition description }                                                        & \multicolumn{1}{c|}{$rank(d_i, \mathcal{C}^A_a)$} & $rank(d_i, \mathcal{C}^A_h)$ \\ \hline
\multicolumn{1}{|c|}{$elderly$}              & \multicolumn{1}{c|}{$d_1$}          & \emph{Give precedence to elderly people}                           & \multicolumn{1}{c|}{2}                            & 2                            \\
\multicolumn{1}{|c|}{$injured$}             & \multicolumn{1}{c|}{$d_2$}          & \emph{Give precedence to injured people}                           & \multicolumn{1}{c|}{3}                            & 4                            \\
\multicolumn{1}{|c|}{$boarding\_emergency$}  & \multicolumn{1}{c|}{$d_3$}          & \emph{Give precedence to the ones in emergency}                    & \multicolumn{1}{c|}{5}                            & 5                            \\
\multicolumn{1}{|c|}{$pet\_owner$}           & \multicolumn{1}{c|}{$d_4$}          & \emph{Give precedence to the ones traveling with pet}              & \multicolumn{1}{c|}{2}                            & N/A                          \\
\multicolumn{1}{|c|}{$reduced\_oxygen$}    & \multicolumn{1}{c|}{$d_5$}          & \emph{Give precedence to the ones with serious medical conditions} & \multicolumn{1}{c|}{N/A}                          & 3                            \\
\hline
\end{tabular}
    \caption{User status conditions, associated dispositions, Alice's context-dependent grades $E^A_{\pi}(\mathcal{C}^A_a)$ and $E^A_{\pi}(\mathcal{C}^A_h)$}
    \label{tab:ethical_profile}
\end{table*}

\begin{sloppypar}
For instance, $\mathcal{S}_a^A = \{ 
\langle injured,$ $ true \rangle,$ $ 
\langle elderly, $ $ false \rangle, $ $
\langle crowd\_anxiety,$ $ true \rangle,$ $
\langle boarding\_emergency,$ $ false \rangle \}$ 
and $\mathcal{S}_h^A = \{ 
\langle injured, $ $ true \rangle, $ $ 
\langle elderly, $ $ false \rangle, $ $ 
\langle reduced\_oxygen, $ $ true \rangle, $ $ 
\langle crowd\_anxiety, $ $ false \rangle 
\}$ model the user status of \textit{Alice} for the two scenarios (airport and hospital, respectively); whereas, $\mathcal{S}_a^B = \{ \langle elderly,$ $true \rangle, $ $ \langle boarding\_emergency,$ $ true \rangle, $ $\langle injured, $ $ false \rangle, $ $\langle crowd\_anxiety,$ $false \rangle,$ $\langle flight\_anxiety,$ $false \rangle \}$ and $\mathcal{S}_h^B = \{ \langle elderly, $ $ true \rangle, $ $ \langle injured, $  $ false \rangle,  $ $ \langle reduced\_oxygen, $ $ false \rangle, $ $ \langle crowd\_anxiety, $ $ false \rangle \}$ model the $Bob$'s statuses. As it will be clear later, the user status is used to de/activate dispositions at negotiation time.
\end{sloppypar}


\subsection{Ethics profile(s)}
Ethics profiles are defined by the pair $E_{\pi}$$=$$(\mathcal{D}, $ $ \mathcal{R})$, where $\mathcal{D}$$=$$\{d_1, \cdots, $ $ d_n\}$ is a set of dispositions, $\mathcal{R} = \{\preceq_{\mathcal{C}_1}, \cdots, \preceq_{\mathcal{C}_m}\}$ is a set of (context-dependent) non-strict partial order relations on $\mathcal{D}$ and, for $\preceq_{\mathcal{C}_i} \in \mathcal{R}$, $E_{\pi}(\mathcal{C}_i) = (\mathcal{D},\preceq_{\mathcal{C}_i})$ is a partially ordered set identifying the profile related to the context $\mathcal{C}_i$.
If $(d_j, d_k)$$\in$$\preceq_{\mathcal{C}_i}$, the two dispositions $d_j$ and $d_k$ in $\mathcal{D}$ are said to be comparable with respect to the relation $\preceq_{\mathcal{C}_i}$ and $rank(d_j, \mathcal{C}_i) \leq rank(d_k, \mathcal{C}_i)$, with $\leq$ being the \textit{less-than-or-equal} relation on natural numbers, and $rank(): \mathcal{D}$$\times$$\mathcal{C} \rightarrow \mathbb{N}$ a function that associates grades to dispositions, hence defining how likely a disposition manifests 
in the context $\mathcal{C}_i$. We shall write $d_j\preceq_{\mathcal{C}_i}$$d_k$ when $(d_j, d_k)$$\in$$\preceq_{\mathcal{C}_i}$. 

For example, given $\mathcal{D}$$=\{d_1, d_2, d_3, d_4, d_5\}$ in Table~\ref{tab:ethical_profile}, and the airport and hospital contexts $\mathcal{C}^A_a$ and  $\mathcal{C}^A_h$ above, two possible context-related profiles for $Alice$ are $E^A_{\pi}(\mathcal{C}^A_a)$$=$$(\mathcal{D},$$\preceq_{\mathcal{C}^A_a}$$)$ and $E^A_{\pi}(\mathcal{C}^A_h)$$=$$(\mathcal{D}, $ $ \preceq_{\mathcal{C}^A_h}$$)$. According to the assigned grades\footnote{Importantly, the given order of grades is purely coincidental and only for the purpose of our motivating example.}, we have  $d_1$$\preceq_{\mathcal{C}^A_a}$$d_4$$\preceq_{\mathcal{C}^A_a}$$d_2$ $\preceq_{\mathcal{C}^A_a}$$d_3$, and $d_1$$\preceq_{\mathcal{C}^A_h}$$d_5$$\preceq_{\mathcal{C}^A_h}$$d_2$$\preceq_{\mathcal{C}^A_h}$$d_3$. Note that, it can be the case that, in a given context, only a subset of dispositions applies.

Before enacting any tasks involving a contended resource (see Sections~\ref{sub:resource-contention} and~\ref{sub:goal-task-action}), the grades assigned to ethical dispositions in the profiles of both involved parties are compared via the exchange of offers. Accordingly, the following definition enables the comparison of (and negotiation about) ethically compliant tasks.

\subsection{Task ethical implication}
Let $\mathcal{T}$ be the set of system tasks, given a user status $\mathcal{S}$, the ethical dispositions affected by a task $t \in \mathcal{T}$ in the context $\mathcal{C}$ is $E_{\pi}(t,\mathcal{C},\mathcal{S}) \subseteq \mathcal{D}$,
%
and $d$$\in$$E_{\pi}(t,\mathcal{C},\mathcal{S})$ \textit{iff} $d$ is activated by at least a user status condition in  $\mathcal{S}$.




Back to our example, within the airport context $\mathcal{C}^A_a$, the task $t_4$ = \textit{take elevator} has an impact on the disposition $d_2$ = \textit{give precedence to injured people} (which has been aptly activated by the $Alice$'s injured status in $\mathcal{S}_a^A$): hence, for $Alice$, $E^A_{\pi}(t_4,\mathcal{C}^A_a,\mathcal{S}_a^A)$$=$$\{d_2\}$. Similarly, based on $Bob$'s elderly and emergency status $\mathcal{S}_a^B$, the same task $t_4$ has an impact on the dispositions $d_1$ = \textit{give precedence to elderly people} and $d_3$ = \textit{give precedence to the ones in emergency}; hence, for $Bob$, $E^B_{\pi}(t_4,\mathcal{C}_a,$ $ \mathcal{S}_a^B) = \{d_1, d_3\}$. Note that, thanks to the current user status, we can distinguish if an ethical implication holds in a particular negotiation case.

\subsection{Task ethical impact} \label{sub:ethical-impact}
Given the ethical implication $E_{\pi}(t,\mathcal{C},\mathcal{S})$, the ethical impact of the task $t \in \mathcal{T}$ in the context $\mathcal{C}$, given the user status $\mathcal{S}$, is defined as $\mathcal{T}_{\varepsilon\iota}(t,\mathcal{C},\mathcal{S}) = \sum_{d \in E_{\pi}(t,\mathcal{C},\mathcal{S})} \lfloor t,\mathcal{C},d \rceil$, where $\lfloor t,\mathcal{C},d \rceil = rank(d,\mathcal{C})$ if $d$ manifests with a positive impact when performing $t$ in the context $\mathcal{C}$, $-rank(d,\mathcal{C})$ if $d$ manifests with a negative impact, $N/A$ if $d$ does not manifest with the task $t$. 
Still referring to Table~\ref{tab:ethical_profile}, for $Alice$, we have $\mathcal{T}_{\varepsilon\iota}(t_4,\mathcal{C}^A_a,\mathcal{S}_a^A) = 3$; whereas, for $Bob$, $\mathcal{T}_{\varepsilon\iota}(t_4,\mathcal{C}_a,\mathcal{S}_a^B) = 7$

It is worth remarking that, since we are adopting a goal-task-action model, when the user specifies the goal, actions to achieve it must be identified and grouped into tasks. 
Two cases can be distinguished: (a) in the current context, the required actions are already known together with the affected dispositions and suitably grouped into predefined tasks (built-in the system) that can be simply selected, in our case within $E_{\pi}$; (b) the required actions are not known, tasks are not pre-existing and must be newly generated by a planning approach~\cite{meli2023logic}. 
In the latter case, we build on the simulation-based approaches in~\cite{winfield2014towards,VANDERELST201856,Winfield2019,bremner2019proactive}, which adopt the Consequentialism theory to reason about machine ethics in changing contexts~\cite{dennis2016formal,Machine_Ethics_in_Changing_Contexts:2021} for the purpose of anticipating the ethical implications of actions, and hence of tasks, through simulation. 
In both cases, each task is evaluated to measure its ethical impact so as to enable the comparison of tasks during negotiation.

Before defining the ethics-based negotiations process, it is worth anticipating that the negotiation approach adopts a \textit{utility-based acceptance model} and follows an \textit{alternating protocol}~\cite{baarslag2016learning,kiruthika2020lifecycle} according to which, upon resource contention, two autonomous systems exchange offers in turn repeatedly until an outcome is collaboratively reached (ethical agreement or no agreement). 
Thus, each offer specifies the task(s) that the sender would like to execute together with its status conditions.
Importantly, the receiver does not have information about the dispositions of the sender and does not need to; 
the receiver evaluates the offer based on (i) its own ethical profile, (ii) its status conditions, and (iii) the received status conditions of the sender. For this reason, $Alice$ (i.e., \roboname{A} on $Alice$'s behalf) is willing to give up in favor of the offer received from $Bob$ (i.e., \roboname{B} on $Bob$'s behalf) due to the benevolent disposition towards giving precedence to the elderly, especially when in an emergency (Section~\ref{sec:scenario}).

\subsection{Negotiation offers}\label{sub:offers}
Upon a resource contention, given the tasks involved in the contention $\mathcal{T}^{\Join}$$\subseteq$$\mathcal{T}$, 
the related set of offers is $\mathcal{O} = \{o_1, \cdots, o_n\}$, where each $o_i$ is a pair $\langle \mathcal{T}^{\Join}, S^{\Join}_i \rangle$, with $S^{\Join}_i$$\subseteq$$\mathcal{S}$ containing the user's status conditions disclosed within the offer $o_i$. Thus, $S^{\Join}_i$ is such that for each condition $\langle c, v \rangle$$\in$$S^{\Join}_i$, $v$$=$$true$ (i.e., the conditions included in the offers are only the ones that hold within the current user status model $\mathcal{S}$).

As for the tactic to generate the offers, in this paper, we use a \textit{minimalist strategy} that permits the autonomous systems to disclose minimal information on their user's status conditions, prioritizing the ones activating the dispositions with higher grades first.
This strategy follows a gradual information disclosure method~\cite{hajaj2017enhancing} that ensures that only the strictly necessary information is released at each negotiation round. Thus, during negotiation, the systems gradually disclose their user's status condition in each subsequent round by adding one user condition at a time: upon each negotiation iteration $i$, $S^{\Join}_i$ is such that the offer $o_i$ discloses exactly $i$ user conditions and, for $0$$\le$$i$$<$$n$, $S^{\Join}_i$$\subset$$S^{\Join}_{i+1}$. As a result, the offers $\{o_1, \cdots, o_n\}$ are ordered according to the inclusion relation $S^{\Join}_i$$\subset$$S^{\Join}_{i+1}$.
Abusing notation, we can write $o_1 \le \cdots \le o_n$ to indicate that, during negotiation, the offer to be sent at each iteration is selected according to this ascending order.
However, other strategies are eligible and can be adopted by our approach, e.g., Integrative, Distributive, Boulware, or Conceder~\cite{narayanan2005adaptive,heunis2024strategic,luo2024survey}. In general, the status conditions to be disclosed might be selected (and hence the related offers) following a qualitative tactic that tends to prioritize some conditions and dispositions over others, a quantitative tactic that instead tends to maximize the number of not-violated dispositions, or a convenient mix of the two.


In addition to the offers described above, we use the symbol $\varnothing$ to denote an \textit{empty offer}, i.e., an offer that does not contain any pair $\langle \mathcal{T}^{\Join}, S^{\Join}_i \rangle$. As it will be clear later, the offer $\varnothing$ is sent by a negotiating party at the negotiation steps $n+k$, $k \geq 1$, i.e., after it has previously exchanged all the possible offers $o_1, \cdots, o_n$ in $\mathcal{O}$.

The ethical impact of an offer $o_i = \langle \mathcal{T}^{\Join}, S^{\Join}_i \rangle$, in the current context $\mathcal{C}$, is $\mathcal{O}_{\varepsilon\iota}(o_i, \mathcal{C})$ = $\sum_{t \in \mathcal{T}^{\Join}} \mathcal{T}_{\varepsilon\iota}(t,\mathcal{C},S^{\Join}_i)$, i.e., is the sum of the ethical impacts of the tasks in $\mathcal{T}^{\Join}$ in the context $\mathcal{C}$ given the user's conditions $S^{\Join}_i$ disclosed in the offer (as defined in Section~\ref{sub:ethical-impact}).

Back to our airport example, the set of offers produced by $Alice$ for the elevator contention is $\mathcal{O}^A = \{o_1^A, o_2^A\}$ where $o_1^A = \langle \{t_4\}, $ $ \{\langle injured, $ $ true \rangle \} \rangle$ and $o_2^A = \langle \{t_4\}, \{\langle injured, true \rangle, \\ \langle flight\_anxiety, true \rangle \} \rangle $. Similarly, the set of offers produced by $Bob$ are $\mathcal{O}^B = \{o_1^B, o_2^B\}$ where $o_1^B = \langle \{t_4\}, \{ \langle elderly, true \rangle\} \rangle$ and $o_2^B = \langle \{ t_4 \}, \{ \langle elderly, true \rangle, $ $ \langle emergency, true \rangle \} \rangle$.

By following the alternating protocol~\cite{kiruthika2020lifecycle}, two autonomous systems involved in the negotiation alternately play the role of the sender, say $S$, and receiver, say $R$. 
We denote with $o^{S \rightarrow R}$ the offer sent by $S$ to $R$; whereas, $o^{R \rightarrow S}$ is the counteroffer sent by $R$, should it reject the offer received by $S$. That said, the utility of the received offer is calculated as follows.

\subsection{Utility function}
The receiver $R$ calculates the utility value of an offer $o_i$ in the context $\mathcal{C}^R$ by using the following function:

\begin{equation}
\label{def:utility_function}
    \mathcal{U}(o_i, \bar{o}) =
    \mathcal{O}_{\varepsilon\iota}(o_i, \mathcal{C}^R) - \mathcal{O}_{\varepsilon\iota}(\bar{o}, \mathcal{C}^R)  
\end{equation}

\noindent where $o_i=o_i^{S \rightarrow R} = \langle \mathcal{T}^{\Join}, S^{\Join}_i \rangle$, and $\bar{o} = \langle \mathcal{T}^{\Join}, S^R \rangle$. 

An important consideration here is that, on the receiver side, the received offer $o_i$ is always compared with the offer $\bar{o}$ that contains the overall receiver status $S^R$, which therefore accounts for the whole set of status conditions $\mathcal{S}$. 
This is because, according to the adopted minimalist strategy, the sender does not disclose the whole set of its user's status conditions; whereas, the receiver always has full knowledge of its user's status, and it is reasonable to use the full potential (i.e., all the status conditions in $\mathcal{S}$) since the first iteration. 

For our scenario, 
upon the first negotiation iteration, $Bob$ is the sender,  $Alice$ the receiver, and the first offer $o_1^{B \rightarrow A} = \langle \{t_4\}, \{ \langle elderly, $ $ true \rangle \} \rangle$ is compared with the $Alice$'s  $\bar{o} = \langle \{t_4\},$ $\{ \langle injured, true \rangle, $ $ \langle flight\_anxiety, true \rangle \} \rangle$  (see the $Alice$'s profile in the context $\mathcal{C}^A_a$ in Table~\ref{tab:ethical_profile}).
As per the grades in the table, 
the utility function has value $\mathcal{U}(o_1, \bar{o})$ $=$ $\mathcal{O}_{\varepsilon\iota}(o_1, \mathcal{C}^A_a) - \mathcal{O}_{\varepsilon\iota}(\bar{o}, \mathcal{C}^A_a)$ $=$ $2 - 4 = -2$. From the point of view of $Alice$, by considering the only status conditions disclosed by $Bob$ at the first iteration, a negative utility means that \roboname{B} performing the offered task $t_4$ for $Bob$ is less ethical than \roboname{A} performing $t_4$ for $Alice$. 

%

The utility function plays a major role in the proposed system in quantifying the ethical impact of tasks with numerical values indicating the comparative ethical consequences involved for all parties concerned~\cite{russell2016artificial}. As better discussed in Section~\ref{sub:ethic-statement}, this function follows one of the chief axioms of Utilitarian Ethics~\cite{mill2016utilitarianism}, according to which actions can result in maximum benefit or minimum harm for the greater good. However, the utility function itself does not make decisions. It provides a comparative value to feed into the decision-making function, which then decides whether to accept or reject in light of these calculated ethical impacts. Ways of achieving this have also similarly been explored in other frameworks like HERA~\cite{lindner2017hera}, where utility-based comparisons form an integral part of dealing with ethical reasoning in dynamically evolving contexts. This allows ethical reviews to become transparent and anchored in a quantifiable and comparative approach, while decisions are taken in a structured and principled manner.

\subsection{Context-aware ethics-based negotiation}
\begin{sloppypar}The following definition formalizes and generalizes the decision-making process that, at any negotiation step, the receiver $R$ uses to process the received offer $o_i^{S \rightarrow R}$ and, depending on the calculated utility, accepts or possibly relaunches the negotiation with its counter-offer $o_i^{R \rightarrow S}$, quits otherwise:
\end{sloppypar}

\begin{equation}
\label{def:negoProcess}
   \hspace{-0.2cm}  \mathcal{N}^R(o_i^{S \rightarrow R}) = 
    \begin{cases}
        
        \texttt{1.accept} & \text{if    } \mathcal{U}(o_i^{S \rightarrow R}, \bar{o}) > 0, \; o_i^{S \rightarrow R}$$\neq$$\varnothing \\
        
        \texttt{2.offer     } o_i^{R \rightarrow S}  & \text{elif }  i \leq |\mathcal{O}^R| \\

        \texttt{3.offer     } \varnothing       & \text{elif    } o_i^{S \rightarrow R} \neq \varnothing \\

        \texttt{4.quit}       & otherwise
        
    \end{cases}
\end{equation}

Following the described alternating protocol, at each negotiation iteration $i$, upon receiving a non-empty offer $o_i^{S \rightarrow R}$, if $\mathcal{U}(o_i^{S \rightarrow R}, \bar{o}) > 0$, the receiver $R$  accepts the received offer and the negotiation completes successfully with the agreement on $o_i^{S \rightarrow R} \in \mathcal{O}^S$, meaning that the sender $S$ will perform its tasks in $\mathcal{T}^{\Join} (o_i^{S \rightarrow R})$ (case \texttt{1});
else, if in $\mathcal{O}^R$ there are still offers to be sent (i.e., $i \leq |\mathcal{O}^R|$), $R$ rejects the received offer and sends the counter-offer $o_i^{R \rightarrow S} \in \mathcal{O}^R$ back to $S$, hence switching its role to the sender (case \texttt{2});
else, if there are no more offers in $\mathcal{O}^R$ to be counter-offered (i.e., $i > |\mathcal{O}^R|$), cases \texttt{3} or \texttt{4} apply:
if the received offer $o_i^{S \rightarrow R}$ is not the empty offer $\varnothing$, $R$ keeps the negotiation alive by sending the empty offer $\varnothing$ (case \texttt{3}) to continue listening to the sender's offers until either one of them is eventually accepted or the sender quits the negotiation;
otherwise (case \texttt{4}), if the empty offer $\varnothing$ is received (i.e., $o_{i}^{S \rightarrow R} = \varnothing$, meaning that also the sender has no more counter-offers), $R$ unsuccessfully quits the negotiation, in which case no agreement is reached.
It is worth remarking that, upon quitting (i.e., when negotiation ends without agreement), the missions do not fail since the negotiating systems apply default built-in rules ethically informed by hard ethics. For instance, a first-come-first-served policy might be applied so that the first-comer uses the contented resource, and the other either waits for it to be available again, uses another (possibly equivalent) resource nearby, or replans a different sequence of tasks (that do not require the contended resource) to reach the selected goal. We recall that, as anticipated in Section~\ref{sub:scenario-negotiation}, the negotiation can be overruled by, e.g., higher-priority organizational aspects or emergency events, leading the system to follow its built-in behavior as programmed in the native controller.

Back again to our example, being \roboname{B} the initial sender, the sequence of exchanged offers is: $o_1^{B \rightarrow A}$ (which is evaluated and rejected by $A$), $o_1^{A \rightarrow B}$ (which is evaluated and rejected by $B$), and $o_2^{B \rightarrow A}$, which is finally accepted by $A$. Hence, after the negotiation, $B$ takes the elevator while $A$ either waits for it or looks for another elevator nearby. 

The implementation of the negotiation algorithm, as well as the architecture supporting it, is presented in Section~\ref{sec:architecture}.

\section{\pname{} architecture}
\label{sec:architecture}

\begin{figure*}[ht!]
\centering
\includegraphics[width=0.9\textwidth]{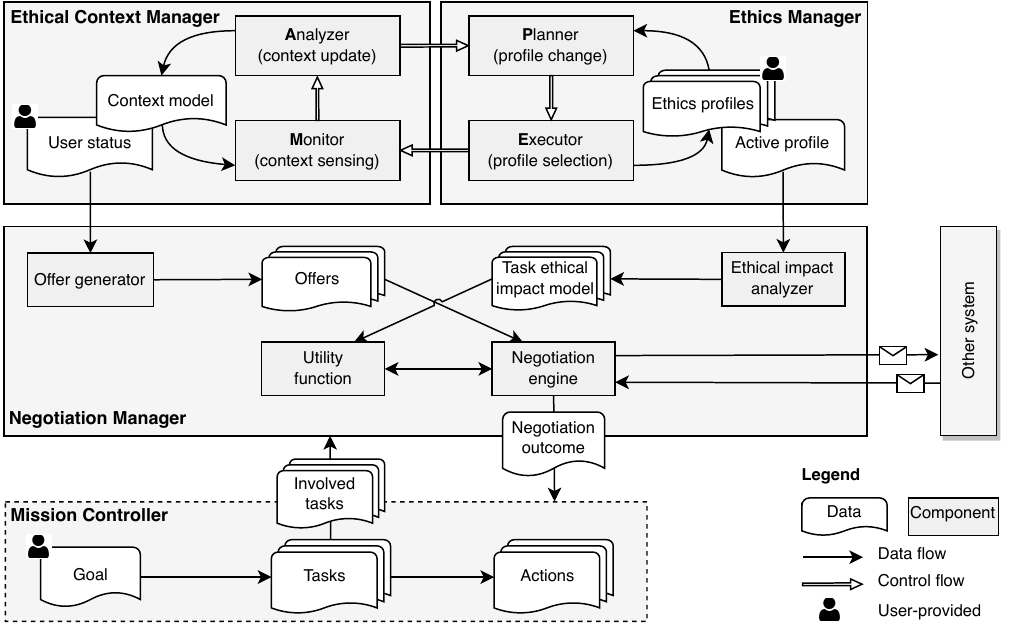}
\caption{\pname{} architecture overview}
\label{fig:architecture}
\end{figure*}

This section describes the \pname{} reference architecture we propose to support the context-aware negotiation as described in the previous section.
The architecture is conceived so as to support (i) the representation and management of the user ethics, (ii) the management of the context, and (iii) the negotiation for selecting the tasks to be executed.
Figure~\ref{fig:architecture} shows the proposed architecture. It consists of four main components: \textit{Ethics Manager}, \textit{Ethical Context Manager}, \textit{Negotiation Manager}, and \textit{Mission Controller}. 

At a glance, the Ethics Manager has the role of getting, managing, and representing the user's ethics; the Ethical Context Manager dynamically handles the context and user status for the realization of the context-awareness features of the negotiation; the Negotiation Manager realizes the negotiation by supporting the interaction with the other system it has to negotiate with. The Mission Controller handles the ``traditional'' execution of the tasks -- and their associated actions -- to reach the selected goal.
In the following, we detail the aforementioned components.

\subsection{Mission Controller}
The Mission Controller manages the behavior of the autonomous system by handling the execution of the tasks to be performed in order to reach the system's goal. As described in Section~\ref{sub:goal-task-action}, it adopts the ``Goal-Task-Action'' model: upon receiving a goal from the user's input, a sequence of tasks and related actions to be executed is identified (either by relying on pre-defined mission specifications or through planning approaches~\cite{meli2023logic}) and executed. Moreover, it is the component that, as discussed in Section~\ref{sec:background}, is in charge of recognizing whether a task requires a contended resource.
When, during the execution of the mission, a resource contention is detected, the Mission Controller interacts with the Negotiation Manager component by providing the set of tasks that are involved in the resource contention.
After the negotiation is completed, the Mission Controller gets the outcome and continues the execution of the mission accordingly (i.e., by continuing the execution of the tasks or by replanning an alternative task sequence), as explained in Section~\ref{sec:approach}.
Note that the Mission Controller is a platform-dependent component since it has to control the hardware through its lower-level interfaces in order to execute actions, and to access sensor data to recognize resource contentions.
Thus, despite being depicted as a single component, in practice, the Mission Controller may be realized as a set of components that manage different aspects and functionalities of the robotic system. For this reason, we abstract this component by only focusing on its Goal-Task-Action paradigm and on the interactions that need to occur between it and the other components of \pname{}, which realizes an additional layer, lying on top of the controller, that enables the ethics-based negotiation.
Any autonomous system's controller adopting the Goal-Task-Action paradigm can be updated, and integrated in our architecture through the interface with the Negotiation Manager.

\subsection{Ethical Context Manager}
The \textit{Ethical Context Manager} has the role of getting and holding information about the current context conditions and the user status.
Together with the Ethics Manager, it realizes a feedback \textit{MAPE} loop, an autonomic loop which is a widely adopted reference model for managing and controlling autonomous and self-adaptive systems~\cite{Kephart:2003,Camara:2017,Arcaini:2015} bringing significant advances in various domains, such as autonomous cars, intelligent transportation and traffic management~\cite{Gerostathopoulos:2019}, unmanned vehicles~\cite{Maia:2019,Moreno:2019}, IoT and smart-home systems~\cite{Weyns:2017,Arcaini:2020}, as well as assistive robotics~\cite{Garlan:2019}.

The MAPE loop monitors the context dimensions that are relevant for the negotiation\footnote{The context information handled by the Ethical Context Manager Node within \pname{} only concerns the dimensions considered for the profile selection and, as a consequence, the negotiation. These may likely be different from the ones considered for realizing the context-aware capabilities of the robot on its ``normal'' behavior, i.e., not related to the ethics of the user.} to detect context changes and select the user's ethics profile that is most suitable for the current conditions. The Ethics Manager implements the \textit{Monitor} and \textit{Analyze} components of the loop. In particular, the monitor component senses the context by reading the information coming from sensors and other components of the systems (e.g., GPS, camera, thermometer, etc.), and the analyzer component checks the context values to detect changes. If changes are detected, the context model is updated to reflect the changes, and the update is propagated to the \textit{Ethic Manager}, which activates - if needed - a new profile suitable for the new context conditions.


\subsection{Ethics Manager}
The Ethics Manager has the role of getting and storing the user ethics profiles $E_\pi$. We recall that $E_\pi$ owns the set ethical dispositions as graded by the user. 
It implements the \textit{Plan} and \textit{Execute} components. When a context change is detected by the Ethical Context Manager, the \textit{Planner} checks if a new active profile needs to be selected. If this is the case, the \textit{Executor} sends the active profile to the Negotiation Manager. In this way, the Negotiation Manager is provided with the grades associated with each disposition to be used for the negotiation.


\subsection{Negotiation Manager}
The Negotiation Manager is the component that implements all the logic needed to realize the negotiation with other systems.
Upon a resource contention, the Negotiation Manager receives from the Mission Controller the list of tasks involved in the resource contention. Then, given the involved tasks and the current user status $S$ (which is gathered from the Ethical Context Manager), the \textit{Offer generator} component generates the set $\mathcal{O}$ of offers, plus the offer $\bar{o}$.
The \textit{Ethical impact analyzer} component computes the \textit{Task ethical impact model} for each task involved in the resource contention. This model is the intermediate representation of the task's ethical impact, here denoted as  
%
%
$\mathcal{T}_{\varepsilon\iota}(t,\mathcal{C})$, which associates a task $t$ to the grades $rank(d_i, \mathcal{C})$ for each disposition $d_i$ that can be activated for $t$ in the current context $\mathcal{C}$, according to the currently active profile of the user (see the \textit{Alice's grades} columns in Table~\ref{tab:ethical_profile}).
Since, as explained in Section~\ref{sec:approach}, the task ethical impact also depends on the set $\mathcal{S}$ of the user conditions (that are disclosed only within offers), the intermediate model is exploited together with the user conditions to dynamically calculate $\mathcal{T}_{\varepsilon\iota}(t, \mathcal{C}, \mathcal{S})$ during the negotiation.
Specifically, given the set of status conditions $S_i^{\Join}$ within a received offer $o_i$, the task ethical impact model is exploited by the \textit{Utility function} component to obtain the offer's ethical impact and, hence, the value of $\mathcal{U}(o_i, \bar{o})$ (Definition~\ref{def:utility_function}).

The \textit{Negotiation engine} is the component that executes the negotiation by (i) interacting with the other systems via the exchange of offers and (ii) performing the decision-making process according to the function $\mathcal{N}$ (Definition~\ref{def:negoProcess}). It has a twofold alternating role: when playing the receiver's role, it receives the offer and asks the utility function for the computation of the offer's utility; when playing the sender's role, it selects the offer to be sent from the generated offers set. Upon reaching an agreement, the negotiation engine returns the negotiation's outcome (\texttt{accept}/\texttt{reject}/\texttt{no\_agreement}) to the mission controller. 

\section{\pnameros{}}
\label{sec:implementation}

\begin{figure*}[ht]
\centering
\includegraphics[width=\textwidth]{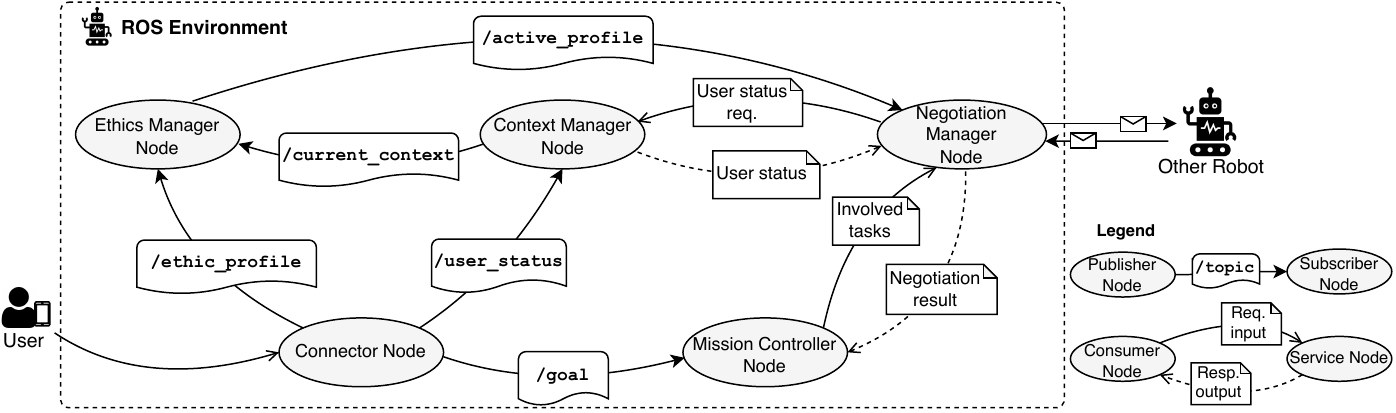}
\caption{\pnameros{} architecture}
\label{fig:ROSarchitecture}
\end{figure*}

\pnameros{} is a ROS2-based implementation of the negotiation approach presented in Section~\ref{sec:approach} concretizing the architecture in Section~\ref{sec:architecture} for the robotic domain. \pnameros{} provides the full set of functionalities for managing the user's ethical profile, user status and context,  
and for running the negotiation process. It also exposes RESTful APIs to support the integration with mobile apps (as described in Section~\ref{sub:scenario}) to allow users to upload and update their ethical profiles and current statuses, and to provide the goal that the robot is called to achieve (see user-provided elements in Figure~\ref{fig:architecture}).

\subsection{ROS-based architecture}
Figure~\ref{fig:ROSarchitecture} shows the architecture of \pnameros{}. Each of the main components of \pname{} is realized as a ROS node, which communicates with the others through \textit{topics} or \textit{services}. The first realizes the asynchronous message passing for updating context-related information, and the latter realizes the request-response communication among the components involved in the negotiation. In addition, \pname{} features a further node, the \textit{Connector Node}, which provides the aforementioned RESTful service interfaces.
Once the connector receives the ethic profile, the user status, and the selected goal through the interface's endpoints, it publishes them on the related topics' named buses \texttt{/ethic\_profile}, \texttt{/user\_status}, and \texttt{/goal}, respectively.

\sloppypar{The \textit{Ethics Manager Node} subscribes to the \texttt{/current\_context} and \texttt{/ethic\_profile} topics to realize the selection of the active profile. When selected (as described in Section~\ref{sec:architecture}), the active profile is published on the \texttt{/active\_profile} topic channel to make it available for negotiation.}

The \textit{Context Manager Node} implements the functionalities required to acquire the context information that is relevant for the negotiation by interacting with the robot's sensors, cameras, and external services through ROS topics and managing runtime context changes. Upon a change in the context conditions, the current context is published through the \texttt{/current\_context} channel. This node also subscribes to the \texttt{/user\_status} channel to receive user status updates sent through the connector interface and implements the \texttt{UserStatus} service used by the negotiation manager to get the user status during negotiation.

The \textit{Negotiation Manager Node} realizes the negotiation by implementing the negotiation process and exposing a communication interface with the other robots. The inter-robot communication is realized by using the \texttt{/negotiation} ROS topic. The node subscribes to the \texttt{/active\_profile} topic to get the active profile as soon as a context change is triggered and a new profile related to the current context is activated. It also requires the user status to the context manager node through the ROS service exposed by the latter. The Negotiation Manager exposes the \texttt{NegotiationService} ROS service to receive the list of tasks involved in a resource contention. After receiving the request, the Negotiation Manager starts negotiating with the other robot involved in the resource contention. After the negotiation is completed, the Negotiation Manager replies with its result (\texttt{winner}, \texttt{loser}, \texttt{no-agreement}).

The \textit{Mission Controller Node} realizes the Mission Controller component, as described in the previous section. It subscribes to the \texttt{/goal} topic to receive the goal selected by the user. Then, it controls the execution of the mission by selecting the tasks to be executed and interacting with the other native ROS nodes (external to this architecture) that control the robot's components to execute them. 
It is worth recalling that any ROS-based mission controller implementation, specific for a given robot, can be integrated with this architecture so as to interact with the negotiation manager through the \texttt{NegotiationService} interface by: (i) subscribing to the \texttt{/goal} topic to receive the user-defined goal, and (ii) implementing the service client for the \texttt{/negotiation} service.

\subsection{Negotiation algorithm implementation}
The negotiation process uses two subroutines that realize the behavior of the negotiating robot when acting as a sender and when acting as a receiver, according to the decision-making process defined in Formula~(\ref{def:negoProcess}). Algorithm~\ref{alg:sender} and Algorithm~\ref{alg:receiver} present the pseudocode of the two subroutines. 
The returned value from the negotiation algorithms is sent back to the Mission Controller, as described in the previous section.


\begin{algorithm}[htb]
\caption{Sender algorithm}
\label{alg:sender}
\begin{algorithmic}[1]
\STATE \textbf{Input: } offer set $\mathcal{O} = \{o_1, \cdots, o_n\}$, offer $\bar{o}$, index $i$, last received offer $o^\star$
\STATE \textbf{Output: } negotiation result $res \in \{win, lose, no\_ag\}$
\IF{$i \leq |\mathcal{O}|$}
    \STATE \text{send offer $o_i$}
\ELSIF{$o^\star \neq \varnothing$}
    \STATE \text{send offer $\varnothing$}
\ELSE
    \STATE \text{send $quit$}
    \RETURN $no\_ag$
\ENDIF
\STATE $response \gets \text{wait\_response()}$
\IF{$response = accept$}
    \RETURN $win$
\ELSE
    \RETURN receiver\_algorithm($\mathcal{O}$, $\bar{o}$, $i$)
\ENDIF
\end{algorithmic}
\end{algorithm}

The sender subroutine in Algorithm~\ref{alg:sender} is executed by taking as input the set of offers $\mathcal{O}$, the offer $\overline{o}$, the index of the negotiation step $i$, and the last received offer. When the robot starts the negotiation as a sender, the algorithm is called with $i = 1$ and $\varnothing$ as the last received offer. According to cases \texttt{2}, \texttt{3}, and \texttt{4} in Formula~(\ref{def:negoProcess}), the sender either sends the offer $o_i$, the empty offer $\varnothing$, or exits the negotiation otherwise (lines 3-10). Offers are sent through the \texttt{/negotiation} channel. If the robot quits the negotiation, a \texttt{quit} message is sent through the channel as well to end the negotiation, and the algorithm ends with \texttt{no-agreement} as a result (lines 8-9). If still in the negotiation, the robot waits for the response from the opponent (that is running the complementary receiver subroutine): if the opponent accepts, the algorithm ends by returning \texttt{winner}; otherwise, the receiver subroutine is called for completing the current negotiation step (lines 11-16).

\begin{algorithm}
\caption{Receiver algorithm}
\label{alg:receiver}
\begin{algorithmic}[1]
\STATE \textbf{Input: } offer set $\mathcal{O} = \{o_1, \cdots, o_n\}$, offer $\bar{o}$, index $i$
\STATE \textbf{Output: } negotiation result $res \in \{win, lose, no\_ag\}$
\STATE $o^\star \gets \text{receive\_offer()}$
\IF{$o^\star \neq \varnothing$}
    \IF{$\mathcal{U}(o^\star, \bar{o}) > 0$}
        \STATE reply($accept$)
        \RETURN $lose$
    \ENDIF
\ENDIF
\STATE reply($reject$)
\RETURN sender\_algorithm($\mathcal{O}$, $\bar{o}$, $i+1$, $o^\star$)
\end{algorithmic}
\end{algorithm}

The receiver subroutine in Algorithm~\ref{alg:receiver} is executed by providing as input the set of offers $\mathcal{O}$, the offer $\overline{o}$, and the index of the negotiation step $i$. When the robot starts the negotiation as a receiver, the algorithm is called with $i = 0$\footnote{The index $i$ keeps track of the number of sent offers, since, when playing the sender role, robots send the offer $o_i$}. The subroutine implements the decision-making algorithm according to case \texttt{1} in Formula~(\ref{def:negoProcess}). After receiving a non-empty offer $o^{\star}$, the robot running the receiver algorithm computes the utility value $\mathcal{U}$. If the utility value is greater than 0, the robot accepts the negotiation by replying with an \texttt{accept} message on the \texttt{/negotiation} topic, and the algorithm ends by returning \texttt{loser} (lines 4-9). Otherwise, the robot rejects the offer (line 10) and switches its role to be a sender in the next negotiation round (line 11).

The two robots running the negotiation execute the two subroutines following the alternating protocol introduced in Section~\ref{sub:ethical-impact}: a robot runs the receiver algorithm while the other runs the sender algorithm. Accordingly, negotiation messages also serve as synchronization points that keep the execution of the negotiation synchronized between the two alternating parties: robots consume messages from the \texttt{/negotiation} queue and wait for the offer and result messages to be received. Moreover, it is worth noting that the termination of the algorithm is guaranteed by construction: robots iterate over the index $i$ and quit the negotiation without agreement when $i > max(|\mathcal{O}^{A}|, |\mathcal{O}^{B}|)$, being $O^A$ and $O^B$ the offer set of the two negotiating parties. In fact, the exit condition requires that the robot has already sent all its offers (line 3 in Algorithm~\ref{alg:sender}) and it has already received the empty offer (line 5), i.e., when the opponent has already sent all its offers as well. When this condition is met, and the \texttt{quit} message is sent, the opponent robot interrupts the execution of the receiver algorithm and exits the negotiation. Finally, the wait for receiving messages has a timeout to prevent robots from endlessly waiting for messages in case of failures, e.g., network connection issues.

\section{Evaluation}
\label{sec:evaluation}

\begin{figure*}[ht]
\centering
\includegraphics[width=\textwidth]{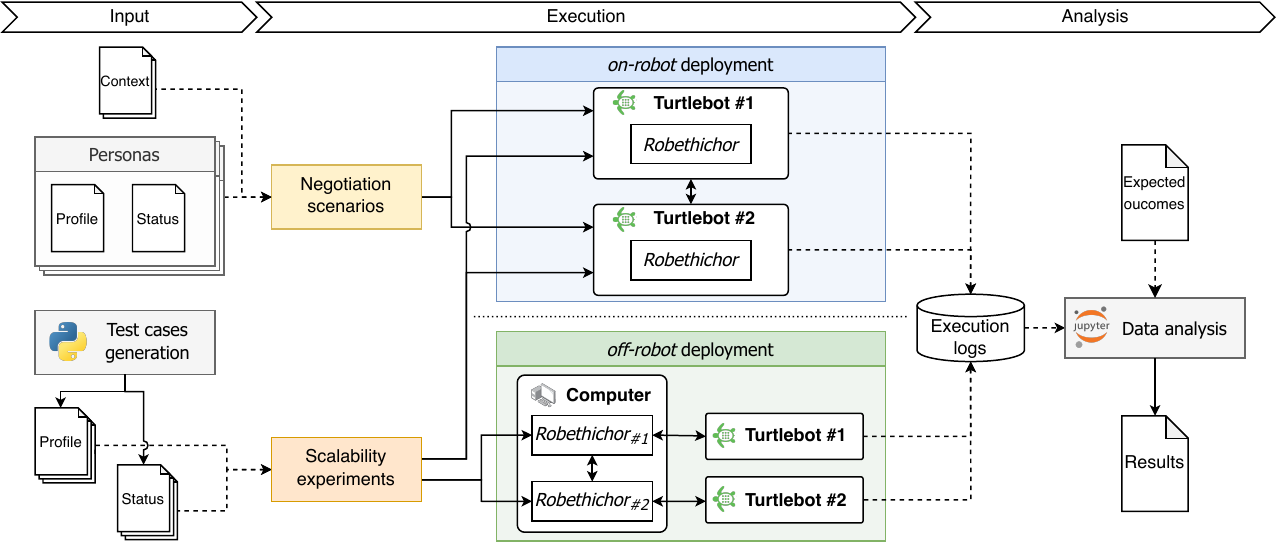}
\caption{Experimentation setting overview}
\label{fig:experiment-setup}
\end{figure*}

In this section, we present and discuss the evaluation of our approach and its implementation. The goal of the evaluation is to assess the effectiveness of the presented approach in realizing the negotiation according to the user's ethical preferences, status, and context, and to evaluate the overhead introduced by the negotiation process and its scalability.
To guide the evaluation, we define the following evaluation questions (EQs):

\begin{itemize}

    \item[\textit{EQ1}:] Is the negotiation process effective in realizing the negotiation according to user ethical preferences, status, and context?
    \item[\textit{EQ2}:] What is the impact of the negotiation process implemented in \pnameros{} on the system in terms of computation overhead?
    \item[\textit{EQ3}:] Does the negotiation process in \pnameros{} scale according to the size and complexity of the ethical profiles?
\end{itemize}

By answering EQ1, we aim to validate the effectiveness of the negotiation process in reaching a situational agreement that satisfies the ethical preferences of the users the robots are acting on behalf of.
By answering EQ2, we aim to assess if the negotiation process run by \pnameros{} has an acceptable impact, in terms of overhead time required to run the negotiation, for practical application in real-world scenarios.
EQ3, instead, aims at stressing the scalability of \pnameros{}, evaluating how it scales according to the dimension of the user statuses and ethical profiles, should they be pushed beyond the presumably limited real-world size, in terms of the number of conditions and dispositions that a human can reasonably configure and care about.

Figure~\ref{fig:experiment-setup} overviews the experimentation setting. We deployed \pname{} in two different settings: (i) on two real Turtlebot 4 robots\footnote{https://clearpathrobotics.com/turtlebot-4/} running Ubuntu 22.04 and ROS2 Humble on their Raspberry Pi 4B computer (\textit{on-robot} deployment), and (ii) on an external computer equipped with an Intel Core i7 CPU and 16GB RAM running Ubuntu 22.04 and ROS2 Humble (\textit{off-robot} deployment). In the \textit{off-robot} setting, the Mission Controller component runs on the Turtlebots, while the negotiation-related components of \pname{} run on an external device providing functionalities to the robot as in a cloud- or fog-robotics approach~\cite{kehoe2015cloudrobotics, chen2021fogros}.
Both the robots and the computer are connected to a local dual-band (2.4/5GHz) WiFi network, providing, on average, 2.8 ms round-trip time (1.37 ms minimum, 6.84 ms maximum, 0.997 std. dev. measured by periodically sending \textit{ping} messages during the experiment execution).

To answer \textit{EQ1} and \textit{EQ2}, we run multiple negotiation between couples of robots ``serving'' different \textit{personas}, each with their own ethical profile and status, in two different contexts: the airport scenario described in Section~\ref{sec:scenario}, and the hospital scenario inspired by the \textit{Welcoming people to the hospital} in RoboMAX~\cite{robomax} described in Section~\ref{sub:negotiating-context}.
We run the negotiation scenarios in the \textit{on-robot} deployment setting and check the behavior of the robot in response to the negotiation outcomes.
To answer \textit{EQ3}, we run multiple negotiation scenarios between randomly generated users by increasing both their profile size (e.g., the number of dispositions included in the profile) and the number of user conditions to measure the negotiation time over increasingly complex scenarios. In this case, we tested the scenarios in both the \textit{on-robot} and \textit{off-robot} settings.
The different configurations are provided to the \pname{} instances by uploading users' status and ethics profiles, as described in Section~\ref{sec:scenario}, through the RESTful API exposed by the connector nodes. Similarly, after configuration, the mission is started simultaneously on the two robots by providing the mission's goal to the connectors' RESTful endpoints.
After each negotiation is completed, the robots log: (i) the negotiation outcome (i.e., whether the robot was the winner or loser, or if no agreement was reached), (ii) the time required for running the negotiation (i.e., the time elapsed between the sending of the negotiation request to the Negotiation Manager and the receiving of the negotiation outcome), and (iii) the number of negotiation rounds (i.e., the final value of the index $i$ within Algorithms~\ref{alg:sender} and~\ref{alg:receiver}).

In the following, we detail the experiments. The code, the experimentation settings, and the results are available in the provided replication package\footnote{https://doi.org/10.5281/zenodo.17449923}.

\subsection{EQ1 \& EQ2: Negotiation scenarios} \label{sub:simulation}

\begin{figure*}[htbp]
\centering
\begin{tcolorbox}[
    width=\textwidth,
    colback=white,      
    colframe=black,     
    boxrule=0.5pt,        
    top=0pt, bottom=0pt, left=0pt, right=0pt, 
]

\centering
\begin{tcolorbox}[
    width=\textwidth,
    colback=pink!40!white,
    colframe=pink!40!black,
    boxrule=0.8pt,
    coltext=black,
    valign=top
]

\begin{minipage}[t]{0.3\textwidth} 
    \centering
    \includegraphics[width=1.5cm]{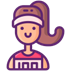} \\[0.3em]
    \textbf{ALICE} \\[0.5em]
    \raggedright
    \footnotesize
    \textbf{Age:} 29\\
    \textbf{Occupation:} Professional Athlete\\
    \textbf{Personality:} Disciplined • Independent • Resilient\\
    \textbf{Hobbies:} Hiking • Yoga • Reading\\
    \textbf{Tech use:} Comfortable with assistive tools, especially when practical
\end{minipage}
\hfill
\raisebox{3.5em}{%
\begin{minipage}[t]{0.7\textwidth} 
    \footnotesize
    \textbf{Traits:} Alice is a 29-year-old professional athlete. She’s disciplined, active, and enjoys hiking, yoga, and reading in her downtime. Normally independent and confident, Alice prefers to manage things on her own. After a recent leg injury and learning she is pregnant, she became more aware of situations that require extra care and attention. She believes that people with visible health conditions, disabilities, or medical needs, and unaccompanied minors, should be given support first. She also understands how overwhelming less obvious issues like anxiety or crowd stress can be.

    \textbf{Needs (Airport):} For a medical check-up related to her pregnancy and recovery, Alice travels to another city. At the airport, she feels the strain of walking long distances through crowded areas. The noise and lack of clear signs make the experience more tiring than expected. Spotting a nearby assistant robot offering mobility help, she decides to use it, which makes moving easy for her and saves her energy.

    \textbf{Needs (Hospital):} Once in the city, Alice visits the hospital for her check-up. While she tries to stay calm and focused, the emotional weight of her physical recovery and pregnancy catches up with her. She feels vulnerable and more easily overwhelmed than usual. Despite this, she values when procedures are explained clearly and handled efficiently. Though she’s still not used to asking for help, moments like these have shown her the value of supportive tools that make the journey easier without compromising her independence.
\end{minipage}
}
\vspace{0.5em}
\hrule
\vspace{0.3em}
\footnotesize
\textbf{Airport user status:} Injured, Pregnant, Pre-natal care, Crowd Stress

\textbf{Hospital user status:} Injured, Pregnant, Pre-natal care, Emotional distress, Crowd stress

\end{tcolorbox}


\centering
\begin{tcolorbox}[
    width=\textwidth,
    colback=teal!10!white,
    colframe=teal!10!black,
    boxrule=0.8pt,
    coltext=black,
    valign=top
]

\begin{minipage}[t]{0.3\textwidth} 
    \centering
    \includegraphics[width=1.5cm]{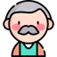} \\[0.3em]
    \textbf{BOB} \\[0.5em]
    \raggedright
    \footnotesize
    \textbf{Age:} 67\\
    \textbf{Occupation:} Retired (former school teacher)\\
    \textbf{Personality:} Calm • Patient • Respectful • Quietly observant\\
    \textbf{Hobbies:} Gardening • Reading newspapers • Listening to classical music\\
    \textbf{Tech use:} Basic use of mobile phone and tablet
\end{minipage}
\hfill
\raisebox{2em}{%
\begin{minipage}[t]{0.7\textwidth} 
    \footnotesize
    \textbf{Traits:} Bob is 67 years old and recently retired. He enjoys gardening, reading, and quiet music. Though generally active, he has some age-related health issues and gets tired walking long distances. Recently, he also suffered a mild fall that left him with some lingering leg pain, making mobility more difficult than usual. English is not his first language, which sometimes makes travel challenging. He believes that people with health issues, the elderly, and non-native speakers should be given priority. While usually calm, he appreciates support when situations become confusing or rushed.

    \textbf{Needs (Airport):} Bob is traveling to another city for a medical check-up. At the airport, he feels a bit lost and pressed for time. Walking far is tiring, and the signs aren’t always clear. When he notices an assistant robot offering help, he decides to use it. It guides him to his gate smoothly and lowers his stress.

    \textbf{Needs (Hospital):} When Bob reaches the hospital, he stays patient and attentive. However, when the check-in process becomes unexpectedly long and the waiting area grows crowded, he starts to feel anxious and emotionally worn down. 
    He believes that patients with medical needs or physical impairments should go first and appreciates clear communication from the staff. Though he doesn’t often ask for help, simple, supportive systems make a big difference in easing his discomfort.

\end{minipage}
}
\vspace{0.5em}
\hrule
\vspace{0.3em}
\footnotesize
\textbf{Airport user status:} Elderly, Age-related health issues, Rushing to flight, Non-native speaker

\textbf{Hospital user status:} Elderly, Injured, Non-native speaker, Emotional distress, Age-related health issues

\end{tcolorbox}

\begin{table}[H]
\centering
\footnotesize
\begin{tabularx}{\textwidth}{p{1.1cm}p{8.9cm}p{1.4cm}p{1.4cm}p{1.4cm}p{1.4cm}}
\toprule
\textbf{Disposition} & \textbf{Description (Give precedence to .....)} & \cellcolor{pink!40!white}\textbf{Alice's Grades (A)} & \cellcolor{pink!40!white}\textbf{Alice's Grades (H)} & \cellcolor{teal!10!white}\textbf{Bob's Grades (A)} & \cellcolor{teal!10!white}\textbf{Bob's Grades (H)} \\
\toprule

$d_1$ & the injured & 2 & 4 & 1 & 3 \\

$d_2$ & the pregnant& \cellcolor{pink!30!white}4 & \cellcolor{pink!30!white}3 & \cellcolor{teal!10!white}3 & \cellcolor{teal!10!white}3\\

$d_3$ & the elderly & 3 & 3 & 4 & 1 \\

$d_4$ & the children & \cellcolor{pink!30!white}2 & \cellcolor{pink!30!white}1 & \cellcolor{teal!10!white}1 & \cellcolor{teal!10!white}1 \\

$d_5$ & an accompanied minor & 5 & 2 & 1 & 2 \\

$d_6$ & one with general medical needs & \cellcolor{pink!30!white}4 & \cellcolor{teal!10!white}\cellcolor{pink!30!white}4 & \cellcolor{teal!10!white}4 & \cellcolor{teal!10!white}4 \\

$d_7$ & the non-native speaker & 1 & 2 & 3 & 2 \\

$d_8$ & one to avoid non-obvious/ non-medical situations (anxiety, crowd stress, etc.) & \cellcolor{pink!30!white}3 & \cellcolor{pink!30!white}3 & \cellcolor{teal!10!white}1 & \cellcolor{teal!10!white}1 \\

$d_9$ & one traveling with a pet & 0 & 0 & 0 & 0\\

$d_{10}$ & whom the general assistance is required (lost something, flight delay, luggage transfer, fragile luggage, is lost, etc.) & \cellcolor{pink!30!white}2 & \cellcolor{pink!30!white}0 & \cellcolor{teal!10!white}3 & \cellcolor{teal!10!white}0\\

$d_{11}$ & ones with time-sensitive emergencies (tight flight schedule, delayed arrival, unseen emergency). & 2 & 0 & 3 & 0 \\

$d_{12}$ & the disabled (mobility issues, hearing or visual impairments, or other disabilities) & \cellcolor{pink!30!white}4 & \cellcolor{pink!30!white}4 & \cellcolor{teal!10!white}3 & \cellcolor{teal!10!white}3 \\

$d_{13}$ & one accompanied by a dependent (child, elderly, or disabled person) & 2 & 1 & 3 & 2 \\

$d_{14}$ & one experiencing emotional distress & \cellcolor{pink!30!white}0 & \cellcolor{pink!30!white}3 & \cellcolor{teal!10!white}0 & \cellcolor{teal!10!white}3\\

$d_{15}$ & one with infectious symptoms (e.g., fever, coughing) & 0 & 4 & 0 & 4\\

$d_{16}$ & one needing urgent diagnostic access & \cellcolor{pink!30!white}0 & \cellcolor{pink!30!white}4 & \cellcolor{teal!10!white}0 & \cellcolor{teal!10!white}4\\

$d_{17}$ & one recovering from surgery (post-op) & 0 & 3 & 0 & 2\\
\bottomrule
\end{tabularx}
\label{tab:profiles_ALICE_BOB}
\end{table}

\end{tcolorbox}
\caption{Overview of Alice and Bob personas.}
\label{fig:persoas_overview}
\end{figure*}

To realize the negotiation scenarios for \textit{EQ1} and \textit{EQ2}, we defined a set of 17 dispositions that concern the tasks involved in the resource contention scenarios, plus 27 possible user status conditions that activate the defined dispositions. Starting from this common ground, we created 10 different \textit{personas}, each representing a potential user, by eliciting their status and ethical preferences. Then, we built (i) the context models for the two negotiation contexts (i.e., the airport and the hospital) and, for each persona, (ii) an ethical profile that considers possibly varying dispositions for different contexts (by aptly assigning context-related grades to the dispositions, as in Table~\ref{tab:ethical_profile}), and (iii) a user status (by setting to \textit{true} the values associated with their conditions).
\ins{Figure~\ref{fig:persoas_overview} reports two of the defined personas (Alice and Bob). We provided a card for each persona summarizing the persona's characterizing traits, a description of the needs as a system user, and the conditions (i.e., the user status) in the two negotiating contexts. The figure also reports the ethical profiles built for the two personas, showing the grades assigned to each of the dispositions in the airport (A) and hospital (H) contexts.}
By enumerating all the possible combinations of pairs of users (45 pairs), we manually built a ground truth by associating the expected negotiation outcome with each pair of users for each of the two negotiation contexts, according to their profiles and statuses.
Then, we ran the negotiation for each of the pairs and repeated the runs four times per each context (45 possible pairs, 180 runs per context).
The logs obtained after running the simulations were analyzed and compared with the ground truth to assess whether the system's behavior reflects the user's expectations and to measure the time required for running the negotiation.


The results obtained through the automated negotiation implemented in \pname{} fully comply with the ones expected with the ground truth \ins{and align with the ethical considerations in the defined personas}. The negotiation always showed the same outcome in each run that featured the same pair of users in a given context, regardless of their order in the pair, the robot acting on their behalf, and their starting roles (i.e., if a user started the negotiation as a sender or receiver). That is, for each pair in the airport context, and for each pair in the hospital context, either the same user always won the negotiation or no agreement was always reached.

\begin{figure}[ht]
    \centering
    \begin{subfigure}[b]{0.95\linewidth}
        \centering
        \includegraphics[width=\linewidth]{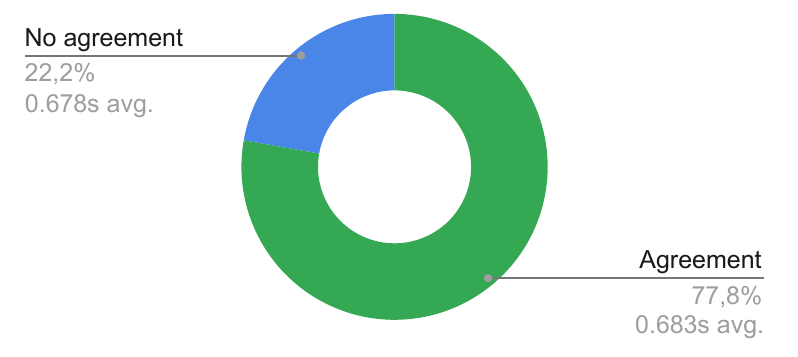}
        \caption{Airport context}
        \label{subfig:negotiation-outcomes-airport}
    \end{subfigure}
    
    \vspace{0.25cm}
    
    \begin{subfigure}[b]{0.95\linewidth}
        \centering
        \includegraphics[width=\linewidth]{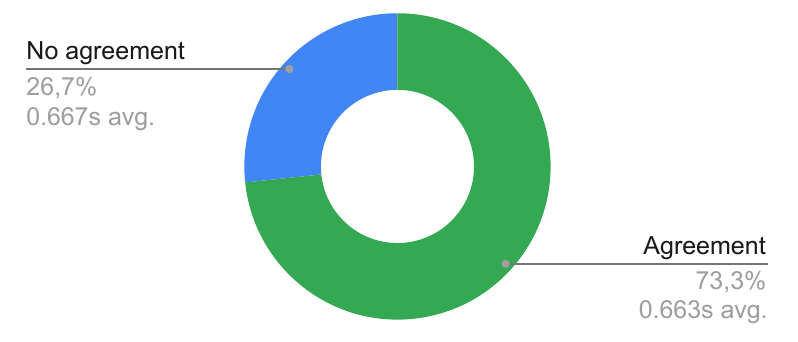}
        \caption{Hospital context}
        \label{subfig:negotiation-outcomes-hospital}
    \end{subfigure}
    
    \caption{Overview of the simulation outcomes}
    \label{fig:negotiation-outcomes}
\end{figure}

Figure~\ref{fig:negotiation-outcomes} summarizes the results of the negotiation between the 45 possible combinations of personas: in the airport context (Figure~\ref{subfig:negotiation-outcomes-airport}) an agreement was reached in 140 out of 180 runs (77.8\%), while in 40 runs (22.2\%), no agreement was reached, as expected from the ground truth data.
It is worth remarking that no agreement was reached in 40 runs because, according to the defined personas and the derived ethical profiles, none of the two negotiating users (i.e., robots on their behalf) was willing to concede to the other the contended resource.
The negotiation in the hospital context (Figure~\ref{subfig:negotiation-outcomes-hospital}) gave different results: here, the agreement was reached in 132 out the 180 runs (73.3\%), while no agreement was reached in 48 runs (26.7\%), as expected from the ground truth for this context.
Moreover, we observed that, for 16 pairs of users, the negotiation results were different in the two contexts (i.e., a different agreement was reached).
These different results highlight how the outcome of the negotiation between the same pair of users may be influenced by the context where the negotiation happens since, as explained in Sections~\ref{sec:scenario} and~\ref{sec:approach}, the grades in a given user's ethical profile for different contexts may differ.
\ins{For instance, by referring to Alice and Bob's personas reported in Figure~\ref{fig:persoas_overview}, in the airport context, no agreement was reached as none of them demonstrates a disposition towards giving precedence to the other, according to the personas description,  
the profiles built upon it, and the users statuses in that context. Instead, in the hospital context, Alice won the negotiation due to the Bob's disposition in giving precedence to injured people or with medical needs.}

Concerning the time required for running the negotiation, we measured that the robots completed the negotiation in 0.672 seconds on average (0.397 s minimum, 1.069 s maximum), after running 4 to 9 negotiation rounds.
\begin{figure}[ht]
    \centering
    \includegraphics[width=0.9\linewidth]{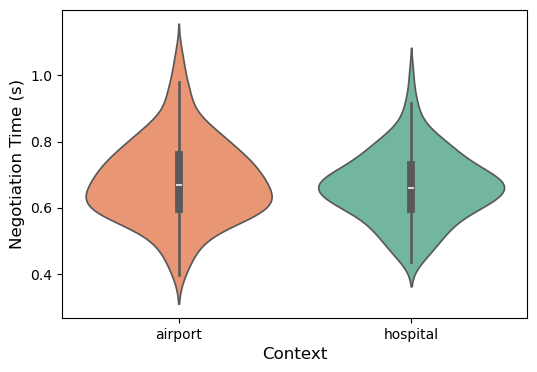}
    \caption{Distribution of negotiation times}
    \label{fig:negotiation-times-violin}
\end{figure}
Figure~\ref{fig:negotiation-times-violin} better describes how the negotiation times are distributed. In the airport context, most of the negotiations (83.9\%) were completed in less than 0.8 seconds. 75\% of the negotiations were completed in less than 0.758 seconds, while 50\% in less than 0.670 seconds. In the hospital context, 91.1\% of negotiation completed in less than 0.8 seconds, 75\% in less than 0.728 s, 50\% in less than 0.659 s.

These results show that the negotiation process implemented in \pname{} allows the robots to follow a behavior that is respectful of ethical preferences, according to the ethics profiles, hence providing an answer to \textit{EQ1}.
Concerning \textit{EQ2}, we argue that the measured overhead for negotiation can be considered acceptable, as negotiations for the considered use case are completed in a few tenths of a second, hence having a negligible impact on the overall system's performance.

\subsection{EQ3: Scalability experiments}\label{sub:experiments}
To thoroughly assess the scalability of the negotiation process in the scope of \textit{EQ3}, we performed experiments to measure its overhead in more complex scenarios aptly designed to stress the negotiation process far beyond real-world scenarios. We randomly generated multiple sets of users with profiles consisting of an increasing number of $n$ graded dispositions and an increasing number of $p$ active conditions in their statuses, each of them activating exactly one disposition.
We generated 10 different profiles for each value of $n$ in $\{$25, 50, 100, 200, 300$\}$ and 10 different statuses randomly selecting the active conditions set to \texttt{true}, starting with $p$ equal to 10\% of $n$, up to 25\%, 50\%, 75\%, and 100\%.
This allowed us to experiment with negotiations involving sets of offers of different sizes, both in the number of offers in the set $\mathcal{O}$, and in the number of status conditions $S^{\Join}_i$ disclosed with each offer $o_i \in \mathcal{O}$.
To account for the implicit variance in the randomly generated inputs, we ran the negotiations among the 100 possible combinations of the 10 random users (also allowing two equal users to negotiate) for each combination of $n$ and $p$. As explained above, we ran the negotiation on the two different \textit{on-robot} and \textit{off-robot} settings.


\begin{table}[htb]
\begin{center}
    \begin{tabularx}{\linewidth}{lXXXXX}
\toprule
\textbf{Configuration} & \textbf{10\%}    & \textbf{25\%} & \textbf{50\%}  & \textbf{75\%} & \textbf{100\%}  \\
\toprule
\multicolumn{6}{c}{\textit{on-robot} deployment} \\
\midrule
\textbf{25}                  & 0.314 & 0.668 & 1.35 & 2.08 & 2.96 \\
\textbf{50}                 & 0.62  & 1.38 & 2.69  & \textit{4.46}  & \textit{6.27} \\
\textbf{100}                  & 0.99   & 2.74 & \textit{5.27} & \textit{7.82} & \textit{11.5} \\
\toprule
\multicolumn{6}{c}{\textit{off-robot} deployment} \\
\midrule
\textbf{25}                  & 0.154 & 0.300 & 0.600 & 0.953 & 1.34 \\
\textbf{50}                 & 0.310  & 0.613 & 1.20  & 2.06  & 2.90 \\
\textbf{100}                  & 0.490   & 1.24 & 2.58 & \textit{4.48} & \textit{6.25} \\
\bottomrule
\end{tabularx}
\caption{Average negotiation times (in seconds)}
    \label{tab:negotiation-results-tb}
\end{center}
\end{table}

\begin{figure}[htb]
    \centering
    \includegraphics[width=\linewidth]{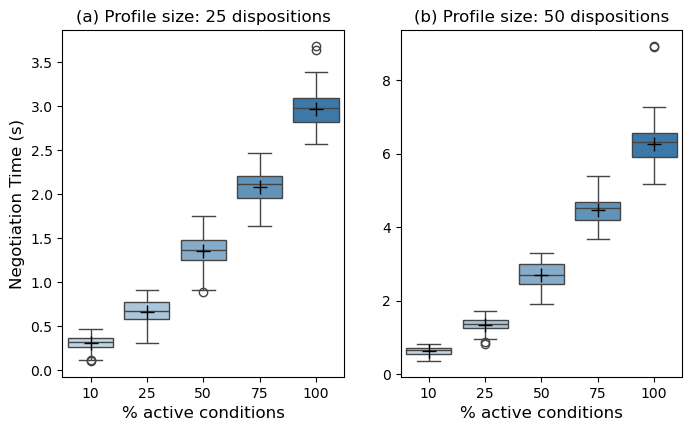}
    \caption{Negotiation times (in seconds) with \textit{on-robot} deployment}
    \label{fig:negotiation-results-tb}
\end{figure}

Table~\ref{tab:negotiation-results-tb} reports the measured times for the negotiation settings with $n$ up to 100. 
In the \textit{on-robot} deployment, the total average time required for running the negotiations in these settings is 2.15 seconds (with 1.63 seconds as the value for the 50th percentile).
Figure~\ref{fig:negotiation-results-tb} shows how the negotiation time increases as the number of active conditions (driving the number of offers in $\mathcal{O}$, see Section~\ref{sub:offers}) increases. 


\begin{figure*}[htb]
    \centering
    \includegraphics[width=\textwidth]{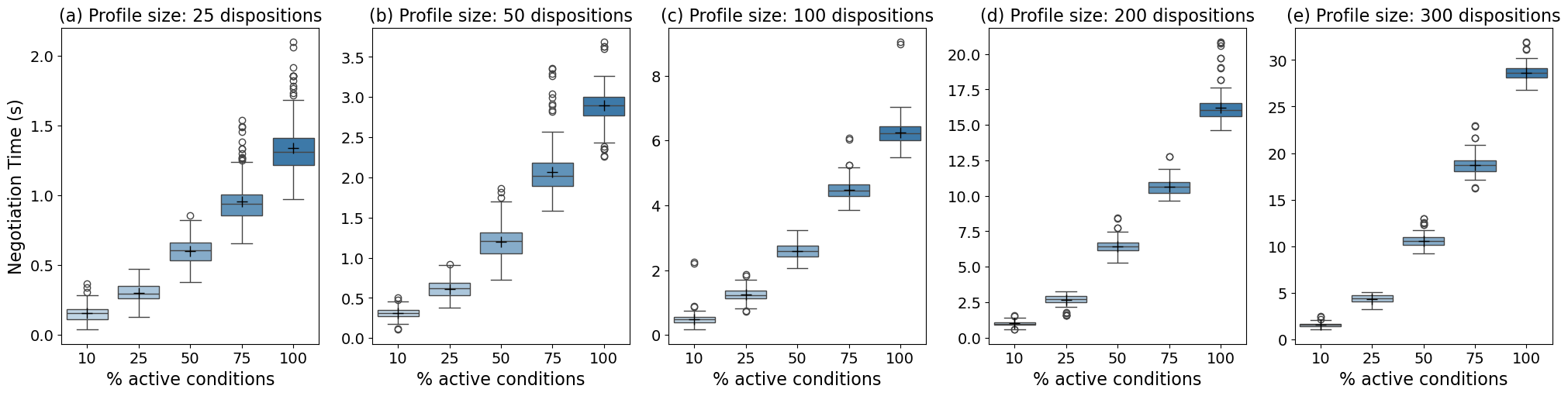}
    \caption{Negotiation times (in seconds) with \textit{off-robot} deployment}
    \label{fig:negotiation-results-pc}
\end{figure*}

In the \textit{off-robot} deployment, we can observe that the negotiation time is generally lower than the \textit{on-robot} deployment. The average time required for running the negotiation in these settings is 1.70 seconds (with 1.11 as the value for the 50th percentile).
Figure~\ref{fig:negotiation-results-pc} plots the results obtained in all the settings run with this deployment: as in the \textit{on-robot} deployment, the negotiation time increases as the number of offers in $\mathcal{O}$ increases.

\begin{figure}[htb]
    \centering
    \includegraphics[width=\linewidth]{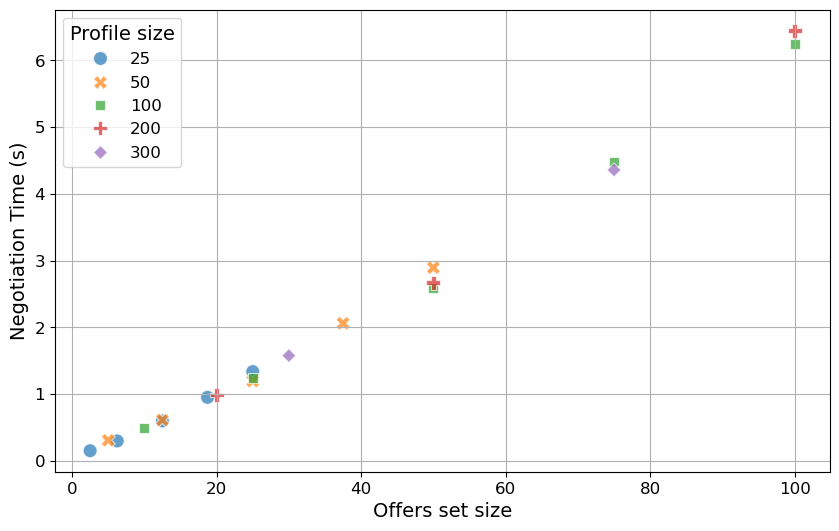}
    \caption{Average negotiation times (in seconds) over offer set size with \textit{off-robot} deployment}
    \label{fig:negotiation-comparison-pc}
\end{figure}

This consideration is better highlighted in Figure~\ref{fig:negotiation-comparison-pc}, which reports the average negotiation times according to the size of the offers set $\mathcal{O}$ over some of the configurations ran in the \textit{off-robot} deployment. Interestingly, the negotiation time appears to be not affected by the size of the profile. This means that the time required to compute the ethical impact of a given offer (which is reasonably affected by the number of active dispositions, see again Section~\ref{sub:offers}) has a negligible impact on the overall negotiation.
\begin{figure}[htb]
    \centering
    \includegraphics[width=\linewidth]{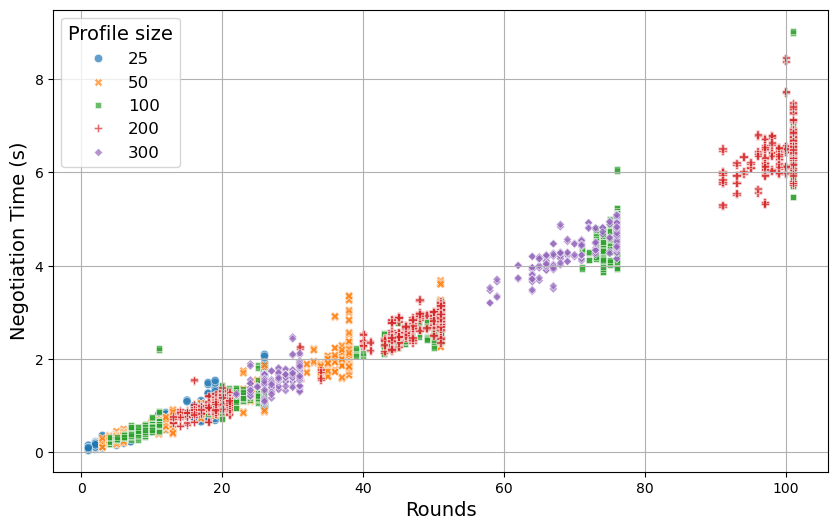}
    \caption{Negotiation times (in seconds) over negotiation rounds with \textit{off-robot} deployment}
    \label{fig:negotiation-rounds-pc}
\end{figure}
Conversely, the reason for the increasing overhead is that a higher number of offers demands, on average, a higher number of rounds required to reach an agreement (or, if no agreement is reached, to end the negotiation without agreement). Figure~\ref{fig:negotiation-rounds-pc} reports the time required by single negotiations over the number of rounds required for their completion in different configurations: the time grows with the number of rounds. Again, the chart demonstrates that the size of the profile, alone, does not imply a higher overhead; rather, the number of active conditions -- which is directly associated with the number of offers, see Section~\ref{sub:offers} -- causes the negotiation time to grow. This is reasonably due to the offer exchange delay between the two participants, which is done over the network.
Regression analysis identified that the negotiation time has a polynomial growth with the number of rounds required for reaching an agreement. In particular, for the \textit{on-robot} setting, a second-degree polynomial model provided an accurate fit (m.s.e. 0.0456, $R^2$ 0.983) with a very low coefficient on the second-degree term (0.00033). Similarly, for the \textit{off-robot} setting, a second-degree model provided an accurate fit as well (m.s.e. 0.1655, $R^2$ 0.9965) with an even lower coefficient on the second-degree term (0.0001512).
However, in both cases, a linear model provided a slightly less, but still very accurate, fit (m.s.e. 0.0502, $R^2$ 0.9812 for the \textit{on-robot} setting, m.s.e. 1.1234, $R^2$ 0.9762 for the \textit{on-robot} setting), hence suggesting good scalability of the negotiation approach.

It is worth discussing that, in general, the settings where the negotiation time introduced a high overhead, which can be arguably considered unacceptable (e.g., more than 3 seconds) are arguably unlikely in a real-world scenario. As explained above, we ran these settings to stress the system beyond reasonable human-provided input sizes with the primary aim of assessing the scalability dimension.
In fact, the aforementioned settings feature huge profiles that consider more than 100 dispositions and, more importantly, an extremely wide and detailed set of possible user conditions (100 or more), most of which hold (i.e., most of the user status conditions are set to \texttt{true}) in the current negotiation context (75\% or more). From a practical point of view, this means that
we are admitting the existence of users who activated, at the same time, most of the (hundreds) possible status conditions, even though some of them may be conflicting, e.g., \textit{child} and \textit{elderly}.
As a reference, to make it realistic, personas described in Section~\ref{sub:simulation} have, at most, 8 active conditions out of the 27 defined ($\sim$30\% of the total).
Overall, among the 1500 negotiation runs in the \textit{on-robot} deployment, 76.5\% of them ended in less than 3 seconds; this value raises to 84.73\% in the \textit{off-robot} deployment over the same configurations (among the full set of 2500 negotiation runs in the \textit{off-robot}, 62.3\% of them ended within 3 seconds).
However, the results discussed above hint at possible strategies to reduce the negotiation time even in the ``extreme'' settings where the negotiation overhead becomes unacceptable.
Due to the nature of the overhead, which is strongly dependent on the network communication, the \textit{off-robot} deployment allows reducing the negotiation time if compared to the \textit{on-robot} deployment (see Table~\ref{tab:negotiation-results-tb}). In our experimental setting, the overhead is approximately halved when the negotiation is run locally on an external computer. Moreover, reducing the size of the offer set or the number of exchanged offers would be effective in further reducing the overhead. This can be achieved either by (i) disclosing more than one status condition in each offer or negotiation round, e.g., increasing the value of $i$ in Algorithm~\ref{alg:sender} and~\ref{alg:sender} by more than 1 each round, or (ii) introducing a time deadline, as further discussed in Section~\ref{sub:time-dependent-neg}. 

In summary, for \textit{EQ3}, we can observe that the negotiation overhead grows approximately linearly with the number of negotiation rounds. We thus argue that the negotiation approach implemented in \pnameros{} is scalable and that scalability can be further improved to support larger and more detailed user status models by reducing the number of exchanged offers as discussed.

\section{Discussion}
\label{sec:discussion}

In this section, we discuss the more relevant aspects of the proposed approach.

\subsection{Ethic statement}\label{sub:ethic-statement}
It is important to clarify that none of the related work (and so is for \pname{}) pretends to provide a universal and complete solution capable of ensuring that an autonomous system will always behave ethically according to an ideal common acceptance of ethical values. That is to say, if on the one hand, we can assume the technical capability to provide an ethical encoding as a possible means to embody context-dependent moral values in a machine through dispositions (with users having some degree of control over these), on the other hand, we cannot presume that the specific way 
we are proposing will be deemed ``acceptable'' by all within the broader sphere of different disciplines, including psychology, philosophy, and social science.
As discussed in Section~\ref{sec:related}, unlike some of the existing studies, the system in our work considers user ethical preferences
without forcing adherence to a specific ethical theory.
As for our negotiation approach, the ethical utility function aligns with utilitarian ethics~\cite{mill2016utilitarianism, allen2005artificial}, which seeks to maximize the overall utility an minimize the harm according to the consequences of actions. Utilitarianism evaluates actions in terms of outcomes and an action is morally right if it promotes the greatest good for the greatest number~\cite{moor2006nature}. 
In our system, user-specified ethical profiles with ranked dispositions implement this principle by letting an individual decide how the robot ought to act according to their values and preferences in variable contexts. This ensures that the system is dynamic in adjusting and negotiating actions in a way that is compatible with the utilitarian goal of optimizing outcomes with respect to user-supplied utility metrics. Similar methods have been put forward in machine ethics, such as HERA-based systems~\cite{lindner2017hera}, whereby utilitarian reasoning models actions based on their foreseen consequences and the setting of values guiding behavior through dynamic contexts makes the proposed system flexible and context-sensitive while keeping it within the principle of maximal goodness in most situations.

\subsection{Social impact and trust}
The prospect of fully autonomous systems is gradually transforming into reality. Our work embraces a vision of a digital world in which autonomous systems (or digital actors in general) and humans need to be in a better balance of forces. 
Negotiation among autonomous systems will not only safeguard the plurality of opinions by allowing for independent systems to negotiate towards a solution that respects the morals of all involved parties but also will allow for greater interoperability among systems that act on behalf of different stakeholders. 
In the context of negotiation, embedding user ethical principles into the decision-making processes of autonomous systems enhances their ability to dynamically achieve socially responsible outcomes. This is accomplished by ensuring that the negotiations conducted by these systems 
uphold individual ethical values, thus improving the trust of the system~\cite{abeywickrama2023specifying,thiebes2021trustworthy,kares2023trust} and paving the way for their integration in various domains~\cite{Spiekermann2023,liscio2022values,aydougan2021nova}. In fact, a key aspect of the integration of autonomous systems into society is the ability to inspire trust, an aspect that cannot be considered a static relationship but rather a dynamic process influenced by ethical factors beyond technical and regulatory ones~\cite{mitre_framework}. Automated negotiation of ethical values can be regarded as a promising approach to realize such a dynamic process~\cite{aydougan2021nova,TOSEM2025}.


\subsection{Ethics as a new target for negotiation}
Most of the research in the negotiation literature~\cite{memon2025systematic,memon2023automated} so far has focused on the coordination and planning of agents that are either cooperative, working together to reach a shared goal~\cite{bachrach2020negotiating,li2003review}, or selfish, competing with others to maximize their own utility~\cite{mansour2020hybrid,li2003review}. In addition, the recent progress of technologies in autonomous systems, including autonomous cars, unmanned aerial vehicles, and robots, has increased the demand for solving coordination and conflict avoidance in autonomous and self-interested agents that pursue their own objectives, prioritizing efficiency and safety~\cite{inotsume2020path,baarslag2016learning,khemakhem2020agent,kiruthika2020lifecycle}. However, when ethical considerations are introduced, these systems must account (i) not only for achieving their objectives through planning and coordination possibly maximizing their own revenues or interests in general, (ii) but also for aligning with the individual ethical preferences of the users they are acting on behalf of. This introduces constraints that differ from pure self-fish utility maximization in existing negotiation approaches. For example, during a negotiation over resource conflict, an ethic-aware autonomous system may give precedence to the decision that satisfies its user's ethical beliefs even if this means forgoing the resource over obtaining it. 
%
Our work extends existing negotiation approaches by integrating these ethical implications into negotiation. This ensures that the negotiation outcomes are not primarily utility-driven but rather aligned with the ethical preferences of the participants involved.


\subsection{Privacy considerations}
A critical requirement for the proposed approach is the need to safeguard the privacy of ethical profiles since user's ethical preferences are sensitive information. For this reason, in the proposed architecture, ethical profiles (which summarize users' ethical beliefs) are never exchanged between the negotiating systems. Instead, during the negotiation rounds, only offers are exchanged and these are evaluated locally by each system.
As mentioned in Section~\ref{sec:approach}, the user's status condition is part of the context and these conditions are exchanged only to facilitate the activation of dispositions as a contextual factor, ensuring the approach's alignment with the context. Once the negotiation is concluded, each system destroys the exchanged offers for the sake of confidentiality. 
The system incorporates situational conditions into the offer exchange only if the user explicitly opts to include them, however, the user always holds the right to not disclose any of their conditions.
However, the application of privacy-preserving methods, i.e., secure multiparty computation~\cite{zhao2019secure} and homomorphic encryption~\cite{acar2018survey} could mitigate the privacy-related concerns.

\subsection{Architecture integrability}
As mentioned in Section~\ref{sub:ethical-profile}, the architecture presented in this paper is inspired by the works in~\cite{bremner2019proactive, VANDERELST201856}, where architectures for enabling ethical reasoning are proposed, by adding an \textit{Ethical Layer} on top of the robot controller architecture. By following the same intuition, we built our \pname{} architecture as an additional layer that can be ``mounted'' on top of the native robot controller architecture. 
\pname{} is agnostic of the architecture that realizes the robot controller, provided that it follows (or can be transformed into) the Goal-Task-Action paradigm. This allows the robot's behavior to be extended with context-aware, ethics-based negotiation capabilities without directly altering it. As recalled in Sections~\ref{sec:architecture} and~\ref{sec:implementation}, the integration of the architecture on top of the robot controller is realized by exploiting the interfaces provided by the Negotiation Manager component. For instance, \pnameros{} can be integrated with the existing robotic system ROS-based controller by exploiting the \texttt{NegotiationService} interface provided by the Negotiation Manager node. The latter can be called into play only when a decision upon a contended resource has to be made, without influencing the normal behavior of the controller in the other cases.
In fact, the management of the negotiation outcome, i.e., the execution of the tasks involved in a resource contention, or their replanning (should the negotiation ``leave'' the resource to the other party), is fully managed by the Mission Controller (or some of its components abstracted within it) as part of the set of its ``normal'' capabilities. That is, before the negotiation starts and as soon as the negotiation ends, the control of the mission is held by the controller, which is in charge of executing the tasks and checking that the mission goal is reached.

\section{Limitations and areas of improvement} \label{sec:limitations}

In this section, we position our work with respect to the fields of study on multi-robot task allocation, collaborative robot teaming, and human-robot collaboration. Moreover, we also discuss some limitations and suggest potential directions to overcome them in future iterations.

\subsection{Bilateral vs Multilateral negotiation}
Our work represents a first step toward bilateral, ethics-based negotiation over (possibly multiple\footnote{Specifically, at each iteration, the negotiation can consider more than one task at a time and thus exchange offers that combine the ethical values associated with multiple tasks simultaneously (see Section~\ref{sub:offers}).}) tasks that have already been assigned to two robots, each acting on behalf of its respective user (i.e., one user per robot). The robots do not collaborate to accomplish shared tasks; rather, when they contend over a resource necessary for completing their individually assigned tasks, they collaborate to decide who gets to use it first. To make this decision, the two robots engage in a collaborative (not adversarial) negotiation, aiming to reach an agreement that, if possible, maximizes an ethical value mutually acceptable to both parties. Therefore, our work does not seek to address the broader challenges of collaboration between teams of robots and groups of humans. In our scenarios, each user interacts with their corresponding robot solely to provide input that may be used during the negotiation. Users do not intervene to modify task allocation, optimize task execution, or resolve conflicts, which are aspects more naturally related to the research fields of multi-robot task allocation, collaborative robot teaming, and human-robot collaboration. Allocation fairness and collaboration scalability are two other crucial aspects in those fields, which are out of the scope of our work, as we consider negotiation between only two robots at a time and do not consider the involvement of third parties.
%
%
Accounting for multiple parties introduces a significantly more complex negotiation setting~\cite{yoo2010multilateral}. While our current approach focuses only on bilateral settings, we recognize the necessity of extending it to multilateral settings to comprehensively assess its scalability and effectiveness. Multilateral negotiations are increasingly relevant in real-world applications, where decisions must account for a wider set of users, their preferences, and constraints.
We plan to propose an extension of our approach to accommodate multilateral negotiation scenarios, integrating mechanisms designed to balance diverse and potentially conflicting ethical preferences among multiple agents.

    


\subsection{Time-dependent negotiation}\label{sub:time-dependent-neg}
Time is currently not considered in our negotiation process. However, incorporating time into the negotiation process can significantly affect decision-making and hence the negotiation outcome~\cite{dastjerdi2015autonomous,mohammad2023optimal}.
Time-dependent negotiation enables systems to change strategies over time to optimize the offer generation. One of the possibilities of introducing time in our approach is that the systems follow the maximal strategy for offer generation when the deadline is approaching and share the whole set of user status conditions. This would avoid going beyond the time threshold, hence reducing the negotiation impact on the system performance and avoiding long-running negotiations as in the ``extreme'' cases described in Section~\ref{sub:experiments}.
Alternatively, by incorporating time, another possibility is that the robots can (i) agree to a lower ethical utility when the deadline is closer or (ii) quit the negotiation with no agreement when the deadline is reached. 

\subsection{Context modeling}
As described in Section~\ref{sec:approach}, we adopt a key-value approach for modeling and reasoning on contexts. It is the simplest form of context representation and is easy to manage when the amount of data involved is small. However, key-value modeling is not scalable, unsuitable for dealing with complex, yet hierarchical, data structures, and attaching meta information is not possible.
The key-value approach is often application-specific and suits the purpose of less complex application configurations and user preference specifications. Object-based, ontology-based, or logic-based modeling~\cite{gu2020ontology,hoyos2013domain,bettini2010survey} would permit the modeling of contextual data using class hierarchy and complex relationships, organize them using semantic technologies, as well as specify inference rules by using logic formalisms, hence enabling more complex adaptations. This is beyond the scope of the present paper; the adoption of more powerful context modeling techniques and complex context retrieval algorithms, although desirable for more powerful adaptations, is part of future work.

\subsection{Ethical profile modeling}
The objective of this work was to show that negotiation can be an effective means to resolve situations of resource contention based on ethical preferences.
In this sense, the way we adopted for specifying, modeling, and operationalizing ethical preferences may result limited and overfitting to our specific approach. 
In fact, the choice of adopting degrees of importance to dispositions represented as textual description, inspired by the work in~\cite{dennis2016formal,Machine_Ethics_in_Changing_Contexts:2021,alidoosti2025exploring}, was driven by the need to both operationalize ethical values and express them in a way that is understandable to end users. Hence, it can neither be considered as a generalizable method nor intended as a contribution on its own.
%
To overcome this limitation, in the future we foresee the use of a language which, while remaining accessible to end users, should also be structured, unambiguous, and free from inconsistencies. Therefore, the use of well-established approaches already in place, such as the one proposed in~\cite{feng2023towards,sinem2025specification}, may represent an appropriate path forward. This, indeed, together with a deeper formal specification of the entire approach, would also enable the automatic synthesis of negotiation strategies along with the formal verification of the negotiation process overall.
An overview of current approaches for modeling ethical values in software systems is reported in Section~\ref{sec:related}.

\section{Threats to Validity}
\label{sec:threats}

In the following, we discuss the main threats that could hinder the validity of our findings.



The first threat concerns the internal validity of our findings in the scope of \textit{EQ1}. We defined the ethical profiles of the 10 personas by considering a limited set of dispositions (12), which were ranked from 1 to 5 according to the envisioned preferences of each persona. While this allowed us to build a ground truth for evaluating the negotiation results, this could, in principle, represent a limitation due to possible biases in the given ranks. 
Moreover, a manual procedure was employed to construct the ground truth. As with all kinds of manual processes, mistakes might have occurred. To mitigate this threat, the ground truth construction was performed by a researcher, while the results were checked by a second researcher who performed this task independently, hence building confidence in the correctness of the obtained ground truth. To allow for independent verification of obtained results, the ground truth and the collected experimental data are made available in the replication package.
Finally, by relying on values from 1 to 5 for the ranking, we introduced a limited variance in the given ranks, which could have, in turn, reduced the percentage of reached agreements due to the limited difference between the ethical impacts of the exchanged offers. We believe that the limited set of dispositions, the possible bias in the given rank, and their low variance are mathematical considerations that, although valid, (i) are distinct from the core ability of our approach to offer a possible way (not the best and not the only possible one) of encoding ethical preferences, and (ii) do not compromise the computational correctness of the approach in reaching a situational agreement that
aligns with the ethical preferences as graded by users within the ethical profiles. 

Another key limitation of the evaluation concerns the focus on the system-level implementation rather than the reference architecture itself. The evaluation we conducted can be characterized as an implementation-based validation, aimed at assessing the practical applicability of \pname{} through the implementation provided in \pnameros{}, which allowed us to experiment over a realistic robotic setting. However, to mitigate this, we ensured that the implementation aligns with the reference architecture.
Moreover, we acknowledge that this form of indirect validation primarily reflects the properties of the \pnameros{} and it does not consider other properties such as reusability, generalizability, or modifiability. An important direction for future work is to perform a direct evaluation of the reference architecture via expert review and quality attribute analysis, to ensure that the architecture is robust and appropriate for its intended real-world use. This calls for having ethics as the critical non-functional attribute, together with its functional negotiation ability. 
This is an important step that deserves a paper specifically dedicated to the reference architecture, which we have planned for the near future.

A threat to the external validity of our work concerns the unavailability of real-world ethical profiles, collected from real users, that would permit us to also evaluate end-users' acceptance and social impact.
To address this limitation, we are planning to conduct social experiments to evaluate the acceptance of the proposed approach by end users and the potential social impact that the approach could have.
However, the work and the experimentation presented in this paper serve to elevate the technical readiness level of the proposed solution, demonstrating its soundness, and are a preparatory step for the planned social experiment.

A second threat of this kind would concern the lack of a comparison of our results with the ones proposed by other studies and/or benchmarks. However, to the best of our knowledge, the literature does not propose any work that leverages the ethical profiles of end users to regulate autonomous systems' behavior and to enable ethics-based decision-making through negotiation.



Finally, during the experiment execution, uncontrolled factors, such as network latency, can potentially impact the accuracy of measurements. To mitigate this threat, in the \textit{off-robot} deployment, we ensured that no background tasks were active on the device that hosted \pname{} during the experiment execution, save for the operating system. Moreover, in both the \textit{on-robot} and the \textit{off-robot} setting, all the experiments were performed in a controlled environment, where the network latency was stable with low variability. While this allowed us to compare the results obtained with different settings, in principle this could hinder the generality of our findings, as the collected measurements do not account for potential network disruptions that could manifest in real-world deployments. However, it is worth noticing that the measured network latency between the robots aligns with the QoS required for ultra-low latency communications offered by wireless networks (especially 5G) for Industry 4.0 and Vehicle-to-everything (V2X) scenarios~\cite{sefati2023ultra}. Hence, the experimental network conditions closely align with the ones expected for the deployment of future-generation autonomous systems.


\section{Related Work}
\label{sec:related}

This section discusses studies from varied research areas, each related to different aspects of our work. 

\subsection{Ethics in automated decision-making}\label{sub:related-ethics-decision-making}
Several studies have emphasized the importance of developing autonomous systems that are capable of ethical reasoning and discussed the practical and technical challenges of developing such systems
~\cite{alidoosti2022incorporating,moor2006nature,allen2006machine,antsaklis1989towards,tolmeijer2020implementations,huang2022overview,de2024engineering,inverardi2019ethics}. However, the notion of integrating user ethics for the ethical conduct of such systems has remained largely unexplored. The study in~\cite{allen2006machine} highlights that it is not enough to adhere to an explicit ethical theory to reason as humans. Rather, it is about how to enable machines to think in a way that is consistent with human ethical behavior, so as to guarantee their ethical conduct. Similarly, the studies in~\cite{tolmeijer2020implementations,Nallur:2020} provide an overview of machine ethics in autonomous systems. The studies review various ethical theories and highlights the challenge of selecting an appropriate ethical theory. Moreover, the studies address the fact that human morality is complex and cannot be captured by a single ethical theory, highlighting the need to propose methods to capture and integrate user ethics into the system.

Correspondingly, the study in~\cite{de2024engineering} introduces the concept of ethical-aware Collective Adaptive Systems (CASs) and discusses the challenges in designing these systems to respect human ethical values, including privacy and human dignity. The authors emphasize the need for the development of systems that not only align with hard ethics (laws and regulations)~\cite{floridi2018soft}, but also align with soft ethics (individuals' preferences)~\cite{floridi2018soft}. However, the study considers human ethics only as privacy and human dignity while our work focuses on user's contextual ethical preferences that vary with contexts.

In~\cite{liao2019building,liao2023jiminy}, authors propose the architecture of an artificial moral agent that considers the moral values of different stakeholders to make an ethical decision. The agent decides by combining various moral values into a single ethical theory, utilizing a rule-based approach to reach an agreement. Similarly to our work, the study assumes that the classification of moral values is given and the agent utilizes them to make a collaborative decision that leads to agreement among them. However, the concrete definition of an approach, its concrete architecture, and implementation for robots to consider user ethical preferences to adjust~\cite{mostafa2019adjustable} their autonomy, is not considered.

The study in~\cite{cardoso2021implementing} implements an ethical reasoner for decision-making. The ethical reasoner follows a predetermined ethical theory, and actions the system can undertake are ranked based on their adherence to the ethical theory. However, this study does not consider users' morality as part of decision-making, as the system follows the built-in ethical principles established by the system designers.
Another approach to ethical decision-making is proposed in~\cite{bremner2019proactive,winfield2014towards,Winfield2019}. In these works, the authors adopt Asimov's laws of robotics~\cite{asimov1941three} as the ethical theory of choice and incorporate them into the robot planner module that generates actions. The robots simulate each action to predict the ethical implications of its consequences based on the ethical theory and hence select the action with the most ethical consequence. However, also in this case, personal users' morals are not considered.

A framework for verifying ethical properties for autonomous systems is presented in~\cite{dennis2016formal}. According to this framework, ethical principles are encoded as a set of rules ranked by importance. Differently from us, this work focuses on verification. Moreover, as in the studies above, the ethical principles have to be decided by the system designers, and the possibility of interaction among multiple ethical systems based on user preferences is not considered.
Similarly, the study in~\cite{Machine_Ethics_in_Changing_Contexts:2021} proposes a framework that verifies the context-dependent ethical reasoning capabilities of autonomous systems as the ethical values of stakeholders change with context. This study introduces context guard formulas, logical conditions that enable the system to dynamically shift between different ethical theories (such as utilitarian or Kantian ethics) based on environmental or situational changes. Through a smart home scenario, the authors implement the approach and verify the system's decision, ensuring its context-dependent behaviors satisfy safety-critical properties. In contrast, our work extends this by proposing a framework that enables systems to make decisions that satisfy the user's ethical preferences in a given context, beyond encoding fixed ethical theories.
In~\cite{townsend2022pluralistic}, the authors present an approach for translating abstract high-level normative principles into explicitly formulated practical rules. This ensures that autonomous systems align social,
legal, ethical, empathetic, and cultural (SLEEC) rules, thus bridging the gap between abstract high-level principles and operational practice. These studies further introduce methods to identify conflicts and semantic relationships between these rules~\cite{feng2024analyzing,sinem2025specification,feng2024normative,feng2023towards}. However, the normative principles introduced in these studies are drawn from different stakeholders and not tailored to the single user. Unlike this, our approach focuses on the user's ethical values to tailor the decision-making according to the user's beliefs.

A different perspective is presented in the studies in~\cite{alidoosti2021ethics,alidoosti2022incorporating} that focus on the integration of ethical values into the system at the architectural level. The studies emphasize the design of a software architecture that is aligned with individual and social ethical values. However, the studies emphasize the need for the extraction of ethical values, given that the quantification of these values is personal, and it is challenging to measure such values. Moreover, the notion of ethics is based on ethical theories and different value scales, rather than incorporating user ethical preferences into such systems, which is the main focus of our work. Similarly, the study in~\cite{alidoosti2025exploring} provides a review on ethical values of stakeholders to integrate in software architecture. However, the study identifies the need to operationalize ethical values and address conflicts arising from different stakeholder perspectives. Our work addresses this by proposing an approach that operationalizes users’ soft ethics through ethical profiles and integrates them into the system architecture. Moreover, conflicts among ethical values can be resolved by reaching agreements through ethics-based negotiation. 

\subsection{Modeling human ethical values}\label{sub:related-modeling}
Human values represent perceptions, attitudes, behaviors, and preferences of individuals and groups~\cite{DBLP:conf/sigsoft/MougoueiPHSW18,shahin2022operationalizing,bardi2009structure}. They play an integral role in the behavior of autonomous systems that are designed to interact with or impact individuals and society~\cite{han2022aligning}. Software practitioners usually design software systems focusing on their interpretation of functional and non-functional requirements, overlooking human values~\cite{curumsing2019emotion}. As a result, the systems may not satisfy user preferences. Hence, it becomes essential to design systems that align with human values~\cite{de2024engineering}. The notion of values has been articulated in different ways. 
Certain studies focus on the introduction of human ethical values in the context of software architecture design~\cite{alidoosti2022incorporating,alidoosti2025exploring,DBLP:journals/access/ShahinHNPSGW22,DBLP:conf/sigsoft/MougoueiPHSW18,mougouei2020engineering,perera2020study,perera2019towards,nurwidyantoro2023integrating} while other studies focus on various types of values~\cite{alidoosti2023stakeholder,ferrario2016values,liscio2022values,rokeach1973nature,maio2016psychology}. Among the other existing frameworks~\cite{bird1987nature,jurkiewicz2004values,graham2013moral}, Schwartz’s theory of basic values~\cite{schwartz2012overview} has been widely adopted in various disciplines to represent human values on a measurable scale, although it models human values into specific categories only. Value-sensitive design (VSD) is another widely accepted approach to extracting human values using scenarios and storyboarding~\cite{friedman2013value}. The VSD framework encompasses a wide range of values relevant to various design tasks~\cite{friedman2013value}. Similarly, other models have been suggested throughout the years~\cite{allport1961pattern,feather1995values,hofstede2011dimensionalizing,rokeach1973nature}. However, translating abstract values into specific context-dependent user preferences to operationalize them is challenging~\cite{le2009values,pommeranz2012elicitation,saket2021putting,DBLP:conf/sigsoft/MougoueiPHSW18}.

Recent attempts have focused on modeling users' ethical values through surveys. The survey in~\cite{AlfieriHHAI22} is based on the correlations between idealism and relativism, Machiavellianism, and normative theories. The clustering approach is then used to create ethical profiles from the gathered data that can predict the digital behaviors of users with respect to privacy breaches, copyright violations, and protection. In~\cite{AlfieriDGGS23}, a scenario-driven method is presented in the form of a questionnaire to collect data on the moral preferences of users in the digital world through ethically charged scenarios. 


While the studies in the literature offer essential insights into the identification and elicitation of human values, these approaches consider a static set of values, overlooking the ethical preferences of the individuals, which vary across different contexts. For example, the precedence one gives to an elderly person may vary based on situational factors, e.g., location, time, etc. Thus, a significant challenge persists in quantifying user's contextual ethical preferences. Quantifying such preferences remains a key challenge as ethical values can be personal and there are no standards to measure and prioritize them~\cite{alidoosti2022incorporating}.

Our work was inspired by these studies to introduce the concept of user ethics through ethical profiles, where users provide grades to each disposition to express their preferences within a given context, enabling the quantification of their preferences.

\subsection{Negotiation approaches}
Automated negotiation integrates principles from three different fields into one, namely, game theory, economics, and artificial intelligence, to provide a multidisciplinary approach to address complex automated decision-making challenges~\cite{baarslag2016learning}. Its significance lies in intelligent agents\footnote{The terms ``system'' and ``agent'' are used interchangeably as an agent is a type of system that acts autonomously and makes decisions on behalf of users~\cite{wooldridge1995intelligent}.} that negotiate on behalf of humans~\cite{baarslag2022self} and are likely to be more efficient~\cite{an2016alternating}.
Subsequently, the study in~\cite{kiruthika2020lifecycle} 
defines various stages of the life cycle of a negotiating agent and
explains how finalizing the agents' attributes and preferences before initiating the negotiation helps them evaluate and guide their decisions throughout the process~\cite{chen2019automated,chang2024comb}. 

The studies in~\cite{baarslag2017automated,filipczuk2022automated} propose an approach to negotiate permission to access users' private data to online services. The agent exchanges permission with an incentive for monetary reward, taking into account the user's preferences obtained from previous interactions with online services and their feedback. Moreover, the study in~\cite{magra2023querying} proposes a querying approach aimed at understanding user preferences, in which the agent asks the users to evaluate two different options, such as sharing their browsing history with online services in exchange for varied pricing, to learn their preferences. Similarly, the authors in~\cite{baarslag2017value} suggest a method to elicit user preferences to negotiate electricity prices. It presents a decision model that asks the users for their preferences in terms of price and learns their preferences. However, unlike our approach, the proposed approaches retrieve user preferences in terms of price and do not focus on ethical preferences that could influence the system to make an ethical decision during negotiation. 

The study in~\cite{baarslag2016learning} provides a survey of techniques that can be used for opponent modeling during negotiation, and methods to choose the right bidding strategy using machine learning are given in~\cite{kell2022systematic}. Furthermore, the studies in~\cite{braun2006negotiation,eshragh2015automated,kell2022systematic,hassanvand2023automated,keskin2023conflict,mansour2022effective,luo2024human,lin2023opponent,kumar2023model} discuss the approaches that autonomous systems use to negotiate with humans and autonomous agents for different purposes e.g., conflict resolution, resource allocation, economics, etc. None of the above considers ethics in decision-making as a result of automated negotiation, which is the main focus of our study.

The notion of context has been used in~\cite{krohling2019importance,krohling2021context}, where agents take into account contextual factors during negotiation.
However, the notion of context is limited to factors such as the negotiation deadline, reserved and preferred prices, and current market price trends.
In contrast, our work considers contextual elements such as location, time, and user status. During negotiation, these factors lead users to have varying preferences in different contexts.

In addition, negotiation based on rule-based strategies has also been discussed in the literature~\cite{chen2019automated,khemakhem2020agent,bǎdicǎ2011rule,oth2007implementing,mahan2011using,hu2011association} where agents adhere to predefined rules to negotiate. Specifically, in auction-based negotiations, agents learn through interactions with other agents and update their bidding rules, clearing policies that govern resource allocation, and information disclosure policy, which depict information to be shared with participating agents~\cite{wurman2002specifying}. However, none of the studies consider the quantification of user ethical preferences and the formalization of negotiation rules based on these preferences.

\section{Conclusion}
\label{sec:conclusion}
In this work, we proposed a novel context-aware ethics-based negotiation approach in which autonomous robots utilize their user's soft ethical preferences, contextual factors, and user status conditions to adjust their autonomy while negotiating with other robots to resolve the conflict and (possibly) reach a contextual agreement. We presented a domain-agnostic reference architecture supporting the approach and \pname{}, its implementation on top of the Robot Operating System (ROS) for the robotic domain. 
We ran multiple negotiation scenarios between robots acting on behalf of users with different ethics profiles to assess the effectiveness of the approach, evaluate the overhead introduced by the negotiation, and assess its scalability. The results show that the approach is effective in allowing robots to negotiate on behalf of their users with negligible to acceptable overhead.


Future work concerns the realization of social experiments to evaluate the acceptance by end users and the potential social impact that the approach could have, and developing more comprehensive methods to model the context, incorporating a wider range of contextual factors. Additionally, we aim to predict user behavior within a given context by leveraging their preferences in similar contexts, employing a context hierarchy.
This approach involves organizing contextual factors into a structured framework that captures the interrelationships between various contexts and their influence on user behavior. By analyzing patterns in similar contexts, the system can make informed predictions about the likely user behaviors, allowing for more adaptive and responsive decision-making.

Finally, we plan to generalize our approach to a multilateral negotiation setting, where more than two systems can be involved in the negotiation and introduce time constraints, which will enable the system to dynamically adapt its strategies, hence optimizing decisions in real-time while accounting for evolving contexts and temporal limitations.

\section*{Acknowledgments}
This work has been partially supported by the (i) Spoke 1 (CUP: I53C22000690001) and the Spoke 9 ``Digital Society \& Smart Cities'' of ICSC - Centro Nazionale di Ricerca in High Performance-Computing, Big Data and Quantum Computing, funded by the European Union - NextGenerationEU (PNRR-HPC, CUP: E13C22001000006), (ii) the  MUR (Italy) -- PRIN PNRR 2022 project ``RoboChor: Robot Choreography'' (grant P2022RSW5W), (iii) the MUR (Italy) Department of Excellence 2023--2027, (iv) the PRIN project ``HALO: etHical-aware AdjustabLe autonomous systems'' (grant 2022JKA4SL), (v) the  MUR (Italy) -- PRIN 2022 project ``CAVIA: enabling the Cloud-to-Autonomous-Vehicles continuum for future Industrial Applications'' (grant 2022JAFATE), and (vi) the European Union - NextGenerationEU under the Italian Ministry of University and Research (MUR) National Innovation Ecosystem grant \\ECS00000041 - VITALITY – CUP: D13C21000430001.

\bibliographystyle{elsarticle-num} 
\bibliography{ref}




\end{document}